\documentclass[
  pre,            % Physical Review E
  amsmath,        % AMS math environments
  amssymb,        % AMS symbols
  reprint,        % Two-column, journal-like layout
  superscriptaddress, % Multiple affiliations — numbered superscripts
]{revtex4-2}
\usepackage{newtxtext}  % Times New Roman-style text font
\usepackage{newtxmath}  % Times-compatible math font
\usepackage{graphicx}   % \includegraphics
\usepackage{dcolumn}    % Align table columns on decimal point
\usepackage{bm}         % Bold math symbols (\bm) — load AFTER newtxmath
\usepackage{comment}
\usepackage{float}
\usepackage[colorlinks=true,           
           linkcolor=red,
            citecolor=red,
            urlcolor=red,
            filecolor=red]{hyperref}   % Clickable references and DOIs — all red

\newcommand{\rv}[1]{\textcolor{black}{#1}}
\newcommand{\id}[0]{\mathrm{i}}
\newcommand{\ed}[0]{\mathrm{e}}
\newcommand{\kb}[0]{\mathbf{k}}

\newcommand{\xb}[0]{\mathbf{x}}
\newcommand{\vs}[0]{\vspace{2mm}}

\begin{document}

% ── Title ───────────────────────────────────────────────────
\title{Numerical simulation of a two-frequency-driven superlattice Faraday-wave pattern}

% ── Authors and affiliations ────────────────────────────────
% REVTeX superscriptaddress style: each \author block is followed
% immediately by its \affiliation(s). Shared affiliations are
% automatically deduplicated into a single superscript by REVTeX.
% Corresponding authors: \email{} goes BEFORE \affiliation{}.

\author{Debashis~Panda}
\affiliation{Department of Chemical Engineering, Imperial College London,
  London SW7~2AZ, United Kingdom}

\author{Nicolas~P\'erinet}
\affiliation{Departamento de F\'isica, Facultad de Ciencia,
  Universidad de Santiago de Chile, Santiago, Chile}
  
\author{Abdullah~M.~Abdal}
\affiliation{Department of Chemical Engineering, Imperial College London,
  London SW7~2AZ, United Kingdom}
\affiliation{Department of Environmental and Sustainability Engineering,
  College of Engineering and Energy, Abdullah Al Salem University,
  12037 Kuwait City, Kuwait}

\author{Lyes~Kahouadji}
\affiliation{Department of Chemical Engineering, Imperial College London,
  London SW7~2AZ, United Kingdom}

\author{Seungwon~Shin}
\affiliation{Department of Mechanical and System Design Engineering,
  Hongik University, Seoul 04066, Republic of Korea}

\author{Jalel~Chergui}
\affiliation{Department of Chemical Engineering, Imperial College London,
  London SW7~2AZ, United Kingdom}

\author{Damir~Juric}
\email{damir.juric@lisn.fr}
\affiliation{Universit\'e Paris-Saclay, Centre National de la Recherche
  Scientifique (CNRS), Laboratoire Interdisciplinaire des Sciences du
  Num\'erique (LISN), 91400 Orsay, France}
\affiliation{Department of Mechanical Engineering, The University of Tokyo, 7-3-1 Hongo, Bunkyo-ku,
Tokyo 113-8656, Japan}

\author{Omar~K.~Matar}
\email{o.matar@imperial.ac.uk}
\affiliation{Department of Chemical Engineering, Imperial College London,
  London SW7~2AZ, United Kingdom}

\author{Laurette~S.~Tuckerman}
\email{laurette.tuckerman@espci.fr}
\affiliation{Physique et M\'ecanique des Milieux H\'et\'erog\`enes,
  CNRS, ESPCI Paris, Universit\'e PSL, Sorbonne Universit\'e,
  Universit\'e Paris Cit\'e, 75005 Paris, France}

% ── Abstract ────────────────────────────────────────────────
\begin{abstract}
The formation of a superlattice pattern in two-frequency-driven Faraday waves discovered and named SSS-I by Arbell \& Fineberg (1998, 2002) 
is investigated by means of Direct Numerical Simulations (DNS) of the full three-dimensional Navier--Stokes equations with a free surface.
Two simulations with distinct quasi-hexagonal initial conditions run at a forcing amplitude $25\%$ above the Faraday-wave onset followed quite different routes, but both led eventually to the same superlattice pattern after around 250 forcing periods.
This regime is inaccessible to the approximations of weak nonlinearity or viscosity. 
The standing-wave pattern contain rows of patches, alternating in time between hills and lakes that are connected by a long skeleton resembing the backbone of DNA strands.
The patches and skeleton of the pattern can be related to its spatial Fourier decomposition,
which combines hexagonal modes with a spatially and temporally subharmonic mode. 
One of the transition routes passes through several fairly long-lived transients including 
different hexagonal patterns and another superlattice pattern; the other passes only through erratic and disordered states.
After another 100 periods, the pattern became unstable and was succeeded by a dynamic version of SSS-I in which the superlattice is modulated and drifts in the direction of the backbone,
while preserving its basic shape.
Convergence to SSS-I states both experimentally in a large geometry and numerically from two different initial conditions and in a minimal geometry demonstrates the robustness of the SSS-I pattern. 
\end{abstract}

% PACS numbers (optional for PRE; may be requested at submission)
% \pacs{05.45.-a, 47.27.Cn, 64.60.-i}

\maketitle
% \linenumbers
\section{Introduction}\label{sec_introduction}

In 2002, \textcite{arbell2002pattern} published a major survey of experimental patterns in Faraday waves,
the standing waves at the surface of a vertically oscillated fluid layer. 
This survey was part of the renaissance in the study of Faraday wave patterns that
was launched in the mid 1980s by 
Gollub \cite{gollub1983symmetry,ciliberto1985chaotic}
and Fauve \cite{douady1988pattern,douady1989oscillatory} and their colleagues,
which was itself part of an explosion of interest in pattern formation that spanned the 1980s to the early 2000s, e.g.\  \cite{golubitsky1988singularities,crawford1991symmetry, cross1993pattern,rabinovich2000dynamics,hoyle2006pattern}.

In classic fundamental articles,  \textcite{Faraday1831forms} first described the eponymous waves scientifically in 1831 and \textcite{benjamin1954stability} calculated their critical forcing frequency and wavenumber for an inviscid fluid by linear stability analysis in 1954.
The Faraday instability is unusual in that its wavelength is determined by the forcing frequency rather than by the depth, facilitating the creation of more varied patterns.
The usual patterns are classic: stripes, squares, and hexagons, but more unusual patterns began to be observed experimentally \cite{gollub1983symmetry, ciliberto1985chaotic}, such as supersquares \cite{douady1990experimental} and quasicrystals \cite{christiansen1992ordered}.
This trend accelerated when Edwards \& Fauve
\cite{edwards1993parametrically,edwards1994patterns}
generalized the Faraday problem
from purely trigonometric oscillation to two-frequency forcing in order to %deliberate 
excite two spatial scales, thus generating a 12-fold quasipattern.

There followed an avalanche of research on Faraday waves forced by both single and multiple-frequency vibration. \textcite{kumar1994parametric}
carried out the first linear stability analysis of the Faraday instability for viscous fluids, followed by \textcite{besson1996two} for two-frequency forcing. 
Experiments using one \cite{binks1997nonlinear}, two
\cite{arbell2000temporally}
and three \cite{ding2006enhanced} frequencies surveyed the parameter dependence of the patterns and revealed new ones, such as triangles \cite{muller1993periodic} and superlattices \cite{kudrolli1998superlattice}.
Zhang \& Vi\~{n}als \cite{zhang1997patterna,zhang1997patternb} derived a quasi-potential equation in the two horizontal coordinates, valid near onset for deep layers of low viscosity, from the three-dimensional Navier-Stokes equations and used it to study pattern selection in the single and two-frequency Faraday configurations. 
Using the Zhang-Vi\~{n}als equation, 
Silber, Skeldon, Porter, and Topaz 
\cite{silber1999parametrically,silber2000two,porter2002broken,topaz2002resonances,porter2004resonant,topaz2004multifrequency} applied the techniques of equivariant bifurcation theory to study the roles played by three-wave interactions, weakly damped modes and the coefficients of the forcing function. 

The limitations of the assumptions of 
weak viscosity and infinite depth were studied by \textcite{skeldon2011scaling} and by \textcite{binks1997effect}, respectively. Chen and co-workers \cite{chen1999amplitude,chen2002nonlinear} derived a weakly nonlinear model valid for finite viscosity, which was extended to two-frequency excitation and finite depth \cite{skeldon2011scaling}. Excellent reviews comparing weakly nonlinear theory to experiment are those of \textcite{westra2003patterns} for the single-frequency case and \textcite{skeldon2015can} for the multiple-frequency case.

\begin{figure*}
    \centering
    \includegraphics[width=0.9\linewidth]
%    {arbell.pdf} 
    {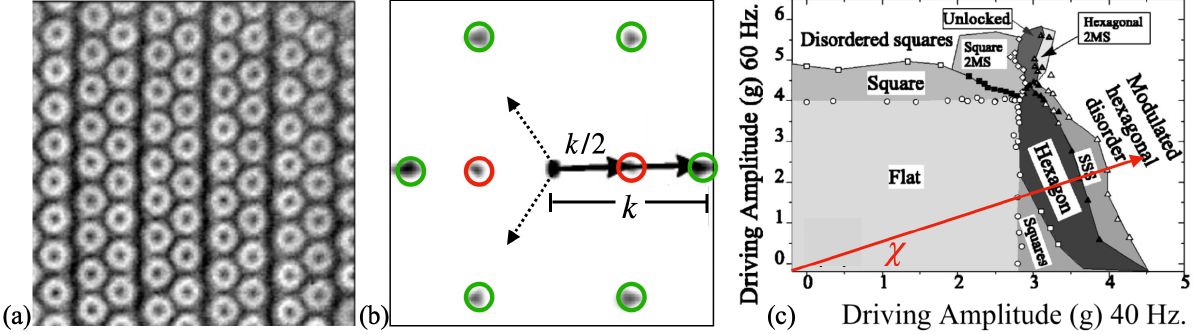} 
    \caption{Experimental SSS-I superlattice pattern, reproduced and adapted with permission from  Figures 4 and 5 of  \textcite{arbell2002pattern}. The density, dynamic viscosity, surface tension, and height of the fluid layer are $\rho_L=950 ~\rm{kg/m^3}$, $\mu_L=0.0219 ~ \rm{Pa.s}$, $\sigma=21.5 \times 10^{-3}~ \rm{N/m}$, $h = 0.155~ \rm{cm}$, respectively, and the imposed vertical oscillation is 
 $F \times[\cos(\chi)~\cos( 2\omega t) + \sin(\chi)~\cos(3\omega t)]$ with fundamental frequency $\omega = 2\pi(20) \rm{rad ~s^{-1}}$ and mixing angle $\chi=30^\circ$. (a) Visualization of height via light intensity. (b) Spatial Fourier transform of (a), showing hexagonal modes (green circles, $k$) and spatially subharmonic modes (red circles, $k/2$). Sometimes patterns are also observed experimentally that contain one or two other wave-vector pairs, indicated here as dotted arrows. (c) Regime diagram showing which patterns appear with different amplitudes $F$ and proportions dictated by $\chi$ of two-frequency forcing with ratio 2:3. The SSS-I state is a secondary bifurcation from hexagons and is observed away from the codimension-two point at which the thresholds corresponding to the frequencies $2\omega$ and $3\omega$ are equal.
}
    \label{fig:Arbell}
\end{figure*}

These studies have focused primarily on pattern selection at or near the onset of Faraday instability.
However, the superlattice states we wish to study occur far from the threshold, where weakly nonlinear theoretical approaches are not suitable. 
Direct Numerical Simulations (DNS) of the Navier-Stokes equations are free from the limitations of weak nonlinearity, as well as any assumptions about depth or viscosity, and can thus be used to study patterns far from onset. We have therefore carried out 
a long-time quantitative study of the dynamic behaviour of 
a two-frequency Faraday configuration. We use the successor of a serial code written by \textcite{perinet2009numerical}, who performed the first three-dimensional DNS of Faraday waves using minimal (single-wavelength) square and hexagonal domains for these two basic patterns. This was followed by the numerical discovery by \textcite{perinet2012alternating} of a complicated and exotic
regime alternating between quasi-hexagons and beaded stripes in a minimal hexagonal domain. 

\textcite{shin2017solver} then wrote a new code, \texttt{BLUE}, based on a hybrid front-tracking/level-set algorithm, which is massively parallelized and thus capable of simulating a much larger domain. \textcite{kahouadji2015numerical} used this code to carry out the first numerical simulation of a superlattice, a supersquare, in which a large square-wave pattern divides spontaneously into sublattices, a pattern that had been previously observed experimentally by Douady and Fauve \cite{douady1988pattern,douady1990experimental}. 
\texttt{BLUE} has since been used to study other related configurations, such as Faraday waves on a sphere \cite{Ebo-Adou_Tuckerman_Shin_Chergui_Juric_2019}, sessile drop vibrations \cite{panda2023axisymmetric}, drop atomisation \citep{PhysRevFluids.9.110514}, and the effect of contaminants on a square-wave pattern \cite{panda2025marangoni}.
A quite different numerical approach was taken in \cite{rojas2010inertial,bestehorn2013nonlinear,richter2019direct}, where lubrication theory was used to simulate thin viscous layers subjected to horizontal and vertical vibrations. Richter and Bestehorn \cite{richter2017thin,richter2021direct} utilised this approach to obtain quasipatterns in two-frequency Faraday waves.

Seeking to match the results of \cite{arbell2002pattern}, 
\textcite{takagi2015numerical} conducted the first numerical simulations of a two-frequency Faraday configuration using a level-set method. 
When volume conservation was assured by adding particles near the interface, they obtained the same squares, hexagons, and rhomboids near the onset of instability as \cite{arbell2002pattern}. However, they were unable to obtain the superlattice patterns they sought, probably because their computational domain was too small or their simulation was insufficiently long; indeed, the limitations of DNS are those of space and time.

Superlattices, first described for Faraday waves by \textcite{kudrolli1998superlattice},
are patterns that have more than one length scale but overall periodicity, in contrast to quasipatterns, which are not spatially periodic \cite{fauve2019quasipatterns}.
This definition is very broad, and indeed, many types of superlattice patterns have been observed, with either single or two-frequency forcing.
Superlattices and quasipatterns can both be formed by combining two hexagonal sets of Fourier modes with the same or different wavenumbers, arranged around a common origin with an angular offset.
Indeed, most superlattices in the literature are of this type, particularly the original SL-I and SL-II patterns of \cite{kudrolli1998superlattice}, the DHS and SSS-II patterns of \cite{arbell2002pattern}, and those called grid patterns in \cite{epstein2006grid}.
Patterns that are not of this category are rhomboids \cite{arbell2000two,arbell2002pattern}, supersquares \cite{douady1990experimental,kahouadji2015numerical}, 2MS \cite{arbell2002pattern}, and SSS-I \cite{arbell2002pattern}. 

Our article presents a detailed numerical simulation of SSS-I, one of the instances of what \textcite{arbell2002pattern} called a 
Subharmonic Superlattice State because the period of its temporal response is twice that of the two-frequency forcing period.
The SSS-I has been studied mathematically by \textcite{rucklidge2003secondary} and by \textcite{Matthews_2004} using the approach developed by \textcite{tse2000spatial}. Aside from this, to our knowledge, there have been no other investigations of the SSS-I pattern and certainly no numerical simulations.

The SSS-I is shown in Figure \ref{fig:Arbell}, which combines and modifies figures from \cite{arbell2002pattern}. 
Figure \ref{fig:Arbell}(a) shows the surface obtained experimentally by light-intensity imaging.
The surface consists of white patches traversed and connected by a dark zig-zagging backbone or skeleton,
somewhat reminiscent of a strand of DNA. The location of the backbone alternates at each forcing period.
Figure \ref{fig:Arbell}(b) shows its spatial Fourier spectrum containing three pairs of  
wavevectors of wavenumber $k$ arranged hexagonally, and another pair of wavenumber $k/2$ which is 
parallel to one of the pairs of hexagonal wavevectors. Dotted arrows of length $k/2$ show wavevectors that are present in other versions of SSS-I which will not be studied here.
Figure \ref{fig:Arbell}(c) shows the regimes obtained in the 2:3 two-frequency-forcing parameter
plane and the direction (mixing angle $\chi=30^\circ$) in the parameter plane that was used both by us and by \cite{arbell2002pattern}
in seeking SSS-I.

We have found the formation of SSS-I to be a long
and complicated process and that obtaining a stable superlattice state from an initial hexagonal state required at least 250 forcing time periods. Moreover, the transition from hexagonal patterns to superlattices is neither unique nor straightforward.
In one of our simulations, the transition proceeds via an erratic sequence of disordered states, while in another, it proceeds via intermediate fairly long-lived states, after which 
it becomes unstable, leading to a new dynamical version of SSS-I in which
the pattern is modulated and drifts along the direction of the backbone.

The remainder of this paper is organized as follows. Section 
\ref{sec:domain} presents the domain chosen for simulating the SSS-I pattern and the governing equations, while section \ref{sec:method} presents the 
numerical method used for solving the equations. Section \ref{ss_validation} presents validations in the two-frequency case, by comparisons with the linear thresholds of \cite{besson1996two} and with the formation of a hexagonal lattice in \cite{arbell2002pattern}, while section 
\ref{sec:fourier} presents the Fourier grid with its relevant modes.  
Section \ref{ss-short} describes the detailed evolution in physical and Fourier space of a simulation carried out at 25\% above the Faraday wave threshold that culminates in an SSS-I superlattice. Section \ref{ss_LONG} describes the results of a longer simulation in which this superlattice loses stability to a dynamic SSS-I state.
We summarize and discuss our results in Section \ref{sec_conclusion}. Appendix \ref{sec:appendix-fourier} presents some mathematical details about Fourier components symmetry, and Appendix \ref{sec:lowerF} describes simulations at lower values of the oscillation amplitude.

\section{Problem formulation}\label{formulation}
\subsection{Domain and governing equations}
\label{sec:domain}
%LSTJun4 Provide F value.
%LSTJun27 Is aspect ratio correct? The hexagons look compressed in y.
%LSTJun27 Domain looks doubled in y
\begin{figure}
    \centering
    \includegraphics[width=1.0\linewidth]{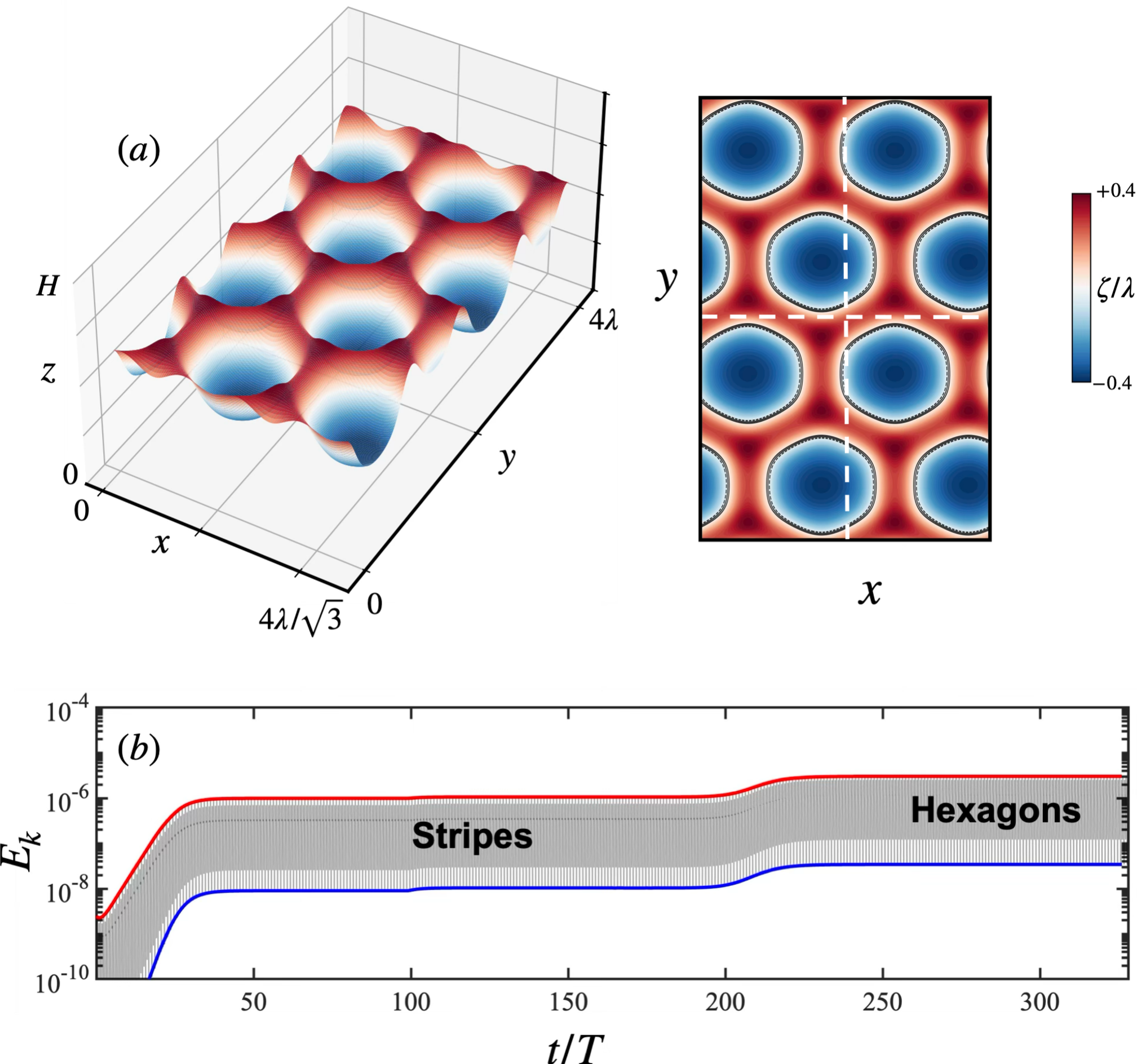}
    \caption{(a) Instantaneous realisation of a hexagonal pattern in Faraday waves in a cuboidal computational box of dimension $4\lambda_c /\sqrt 3 \times 4 \lambda_c \times H$. Left: full three-dimensional domain. Right: two-dimensional projection of interfacial height $\zeta$. 
    The domain used to simulate SSS-I is twice as large in each direction as the minimum domain for accomodating a hexagonal pattern, indicated by white dashed lines. 
    The density and viscosity in the liquid phase (air) below (above) the free surface are $\rho_l$,  $\mu_l$ and $\rho_a$, $\mu_a$. 
    %LStAug15 Is this still true?
    %The domain is doubled in the $x$ direction for the sake of visualisation.
    (b) Temporal evolution of the system's kinetic energy. The gray region represents the evolution of kinetic energy, while the red and blue envelopes represent the maximum and minimum kinetic energy over a window of one time period.}
    \label{fig-1}
\end{figure}
Our three-dimensional domain, shown in figure \ref{fig-1}, is bounded by $z = 0$ and $z =  H$, and 
for our superlattice simulations, has a lateral area 
$ 4\lambda_c/\sqrt 3 \times  4\lambda_c$, where  $\lambda_c$ is the critical wavelength at the onset of the Faraday instability, 
and the ratio of $\sqrt{3}$ between the two sides is required in order to accommodate a hexagonal pattern in a periodic rectangular domain \cite{perinet2009numerical}.
%LSTAug15
Note that the lateral directions used here are each twice as long as those used to simulate a simple hexagonal pattern. These dimensions are necessary to accommodate the SSS-I pattern, and we will see in section \S \ref{ss-short} that even these lengths, while adequate, may be insufficient to capture all effects.
%LSTJun13

The domain consists of air and liquid phases, both Newtonian, incompressible, and immiscible fluids, with densities $\rho_a$ and $ \rho_l$, and dynamic viscosities $\mu_a$ and $ \mu_l$, respectively. 
Initially, the two fluids 
form horizontal layers separated by an interface with surface tension $\sigma$. The height of the lower fluid is $h$. 
%The simulations we wish to study are in the reference frame of the container to satisfy 
We impose 
no-slip and no-penetration boundaries at $ z = 0$ and $ z =  H$, while the boundary conditions at the lateral sides are periodic.
We write the equations in a frame of reference that oscillates with the imposed two-frequency acceleration
\begin{equation}
    F\left[\cos (\chi)~\cos ( k\omega t) + \sin(\chi)~\cos(l \omega t + \phi)\right], 
    \label{eq:accel}
\end{equation}
where $k$ and $l$ are co-prime integers that each multiply the fundamental frequency $\omega$, $\chi$ is the mixing angle that determines the proportions of the two frequencies, 
and $\phi$ is the phase angle between the two waves.

The governing equations 
%for the incompressible fluids 
are given by
\begin{subequations}\label{eq-gEq}
\begin{align}
    \nabla \cdot {\mathbf u} &= 0, \\ 
     \rho \dfrac{\rv{\rm D}{\mathbf u}}{\rv{\rm D} t} &= - \nabla  p+ \nabla \cdot (2 \mu{\mathbf D}) 
    +\int _{ A(t)} \sigma \kappa \,\mathbf n \,\delta({\mathbf x}-{\mathbf x_f})~d A \nonumber \\
    &- \rho g ~\left[1- F(\cos (\chi)\cos(k \omega t) + \sin(\chi) \cos(l \omega t)\right]\mathbf e_z,
\end{align}
\end{subequations}
where \rv{$\rm D$} is the material derivative, $\mathbf{u}$ represents the velocity, $ {\mathbf D} = ( \nabla {\mathbf u} + ( \nabla {\mathbf u})^{\rm T})/2$ is the rate-of-deformation tensor, and $p$ denotes the pressure. 
 The interface is defined by the three-dimensional Dirac-delta function $\delta({\mathbf x} -  {\mathbf x}_f)$, where ${\mathbf x}_f = (x,  y,  \zeta( x, y, t))$ is the interface position at any time $t$, and $\mathbf{e_z}$ is the unit vector in the vertical direction.
 The interface has mean curvature $\kappa$ and unit normal  $\mathbf n$ %to the interface 
 (directed upward to the lighter fluid phase); $F$ is the amplitude of the oscillatory acceleration scaled by the gravitational acceleration $g$, and $A$ is the surface area defined by the instantaneous position of the interface. 

 In our one-fluid formulation, the local density and viscosity %material properties %$\left(\rho\left(\mathbf x, t\right), \mu\left(\mathbf x, t\right)\right)$ 
 are given by
\begin{subequations}
    \begin{gather}
        \rho (\mathbf x, t) = \rho _a + (\rho_l - \rho_a)\mathcal H(\mathbf x, t), \label{eq:rho}\\ 
        \mu (\mathbf x, t) = \mu _a + (\mu_l - \mu_a)\mathcal H(\mathbf x, t)
    \end{gather}
\end{subequations}
where $\mathcal H$ is a smooth Heaviside function, fixed to $1$ for the liquid phase and $0$ for the air. The required boundary conditions for $\mathcal H$ are periodicity on the lateral sides and Neumann conditions at the top and bottom. 
%
%LSTJun13 Is there a coupling equation in addition to mass conservation?
%A coupling equationis required  to complete the system of equations.
Mass conservation of incompressible fluids 
is expressed by 
$D \rho(\mathbf x, t)/ D  t = 0$, which, from Eq. \eqref{eq:rho}, leads %. In the view of discontinuous phases, 
%this reduces 
to $D \mathcal H/ D t = 0$. 
Therefore, the interface is implicitly represented by the Heaviside function and advected by the material motion of the fluids.
%
%LSTJun13 stated previously
%Initially, the interface is flat (unless stated) and the fluid is motionless.
%velocity in the domain is $0$. 
%The pressure field is distributed according to the vibratory volumetric force and depth of the liquid. 
%LSTJun12 BCs at top and bottom previously stated, moving up other bcs.
%The velocity boundary conditions on the lateral sides are periodic, while no-slip and no-penetration is implemented on the top and bottom boundaries. 
%Boundary conditions for the Heaviside function are also required, for which we implemented periodicity on the lateral sides and Neumann conditions at the top and bottom.    
%
\subsection{Numerical Method}
\label{sec:method}
The computational domain is discretised using a uniform three-dimensional finite-difference grid. This grid allows a standard staggered Marker-And-Cell (MAC) arrangement \citep{harlow1965numerical}, where velocity components are stored at the cell faces, while scalar variables, such as pressure, are located at the cell centres. Each grid cell has a dimension $\Delta x \times \Delta y \times \Delta z$. Within this domain, the two fluids are separated by a two dimensional interface, which is discretised by triangular elements. 

Once the initial and boundary conditions are specified, the velocity and pressure fields are computed using a  projection method. Time integration is performed with a second-order Gear scheme.
For spatial discretisation, a second-order scheme is used for scalar variables, while the nonlinear advection terms are handled using an essentially non-oscillatory (ENO) method \citep{sussman1994level}.

The capillary forces (per unit volume) are computed using a hybrid level-set/front-tracking method \citep{shin2009hybrid}. 
The information exchange between the fixed grids and the deformable interface is governed by the immersed boundary method \citep{peskin2002immersed}. The interface, represented by the triangular elements, is advected according to $d\mathbf x_f/ dt = \mathbf u(\mathbf x_f)$, where $\mathbf u(\mathbf x_f)$ is the interpolated velocity at the interface position obtained from the fixed grids' cell faces. 
%LSTJun13 Above, said Gear scheme -- is Gear for the bulk and R-K for the interrace?
The integration is 
%LSTJun13 "achieve" is for a goal
%achieved 
performed
using a second-order Runge-Kutta method. The adaptive time step is chosen as the minimum of the time scales of the
 viscous, capillary, interfacial, advective, and vibratory evolution.   

The 
%LSTJun13 To me "numerical method" means Gear, Runge-Kutta, finite-differences, etc.
%numerical method 
code
is parallelised 
%LSTJun13
%using domain decomposition,
over subdomains each 
containing $32^3$ grid points.
The velocity field is solved using a parallel Generalised Minimal Residual (GMRES) method, while the pressure Poisson equation is solved by a parallel multigrid method. Communication between the parallel processes is managed using Message Passing Interface (MPI) routines. More details of the numerical framework can be found in \cite{kahouadji2015numerical, shin2002modeling,shin2017solver}.

\subsection{Model assessment: linear theory, rolls, 
and hexagons}\label{ss_validation}
To validate our numerical method, we compare the threshold acceleration to destabilise an interface to that of the two-frequency Floquet linear stability analysis of \textcite{besson1996two}. 
For this test, the computational domain is $\lambda_c/2 \times 2\lambda_c \times H$, 
%LSTJun13 already stated
%where $\lambda_c$ is wavelength. Here, 
where here 
$H$ is $8 ~ \rm{mm}$. The mesh resolution is $\Delta x = \Delta y = \lambda/128, \Delta z = H/128$. The domain is decomposed into $2 \times 8 \times 4$ subdomains of each $32^3$ grids, such that the global resolution is 
$64 \times 256 \times 128$ grid points. The 
lower fluid is silicone oil of density $\rho_l = 950 ~\rm {kg/m^3}$, dynamic viscosity $\mu_l = 0.02 ~\rm{Pa . s}$, and surface tension, $\sigma = 20.6 \times 10^{-3}~\rm{N/m}$.
and occupies 
height $h = 3 ~\rm {mm}$. The lighter fluid is air of density 
$\rho_a = 1.206 ~\rm{kg/m^3}$ and viscosity $\mu_a = 1.82 \times 10^{-5} ~\rm{Pa.s}$. 
The imposed two-frequency acceleration \eqref{eq:accel}
has fundamental angular frequency of vibration $\omega = 2 \pi \times 11~ \rm{rad/s}$ and $k = 4$ and $l = 5$. The phase difference $\phi$ is $0^\circ$ and the mixing angle $\chi$ is $45^\circ$. 

\begin{figure}
    \centering    \includegraphics[width=\linewidth]{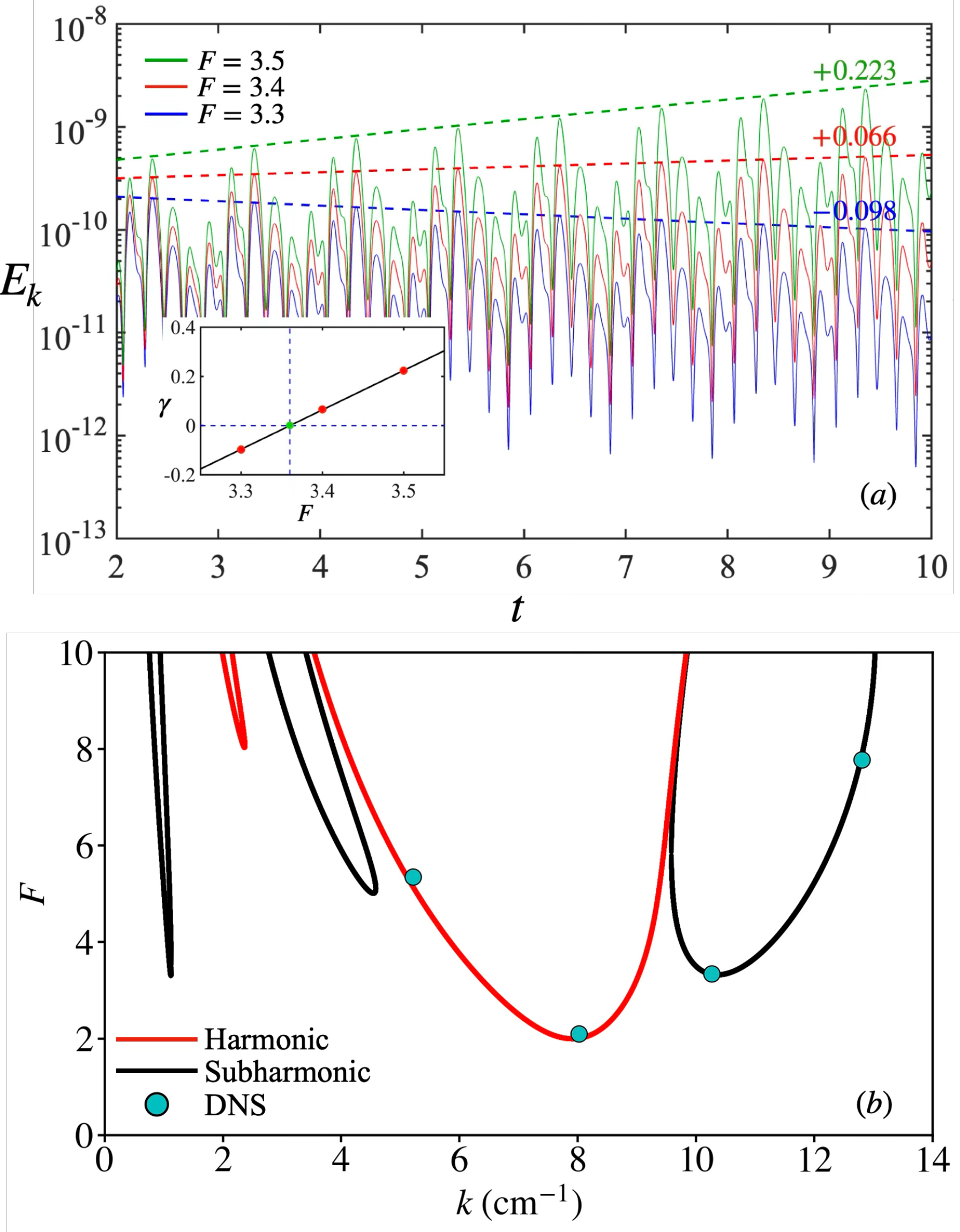}
    \caption{
    Validation by comparison with configuration in
    \textcite{besson1996two}.
The physical parameters are
$\rho_l = 950~\rm{kg/m^3}$, $\mu_l = 0.02~\rm{Pa.s}$, and $\sigma = 20.6\times 10^{-3}~\rm{N/m}$. Forcing parameters are $\omega/2\pi = 11 \rm{Hz}$, $(k,l) = (4,5)$, and $\chi = 45^\circ$.
    (a) Kinetic energy 
    evolution 
    for $F = 3.3, 3.4, $ and $3.5$, 
    with the envelope shown as a dashed line.
The growth 
rates for the three values of $F$ 
are plotted in the inset and a linear interpolation is shown to describe the method of obtaining the threshold acceleration. 
%LSTJun12 Other param values, like $k$?
(b) The DNS results for the thresholds (dots) are compared to the boundaries of the harmonic and subharmonic instability tongues obtained by the linear stability analysis 
of \cite{besson1996two}.
 The thresholds for the two tongues are equal at the bicritical or codimension-two point, which is $\chi=60^\circ$
 for this configuration.
}
    \label{fig-2}
\end{figure}
To determine the 
computational 
threshold acceleration, we ran multiple simulations of acceleration amplitude 
close 
to the critical acceleration. One such example is summarised in figure \ref{fig-2}(a). First, we set the critical wavelength, $\lambda = 0.6125 ~\rm{cm}$ 
($k=10.258~\rm{cm}^{-1}$), and ran 
cases with 
$F = 3.3, 3.4, $ and $3.5$.  
We then extract the envelope of the maximum kinetic energy over each period.
The growth rate $\gamma$ 
is evaluated as the slope of the kinetic energy envelope in a semi-log plot. 
We obtained $\gamma = -0.098, 0.066,$ and $0.223$ for the 
acceleration amplitudes $F=3.3$, $3.4$, and 3.5, respectively.
Assuming that the growth rate at the onset of instability depends linearly on the control parameter $F$, we use linear interpolation to $\gamma=0$ to 
compute the threshold acceleration. This process is summarised in the inset plot of figure \ref{fig-2}(a). 
The same procedure is used to evaluate the threshold acceleration at three other 
wavenumbers, as shown in figure \ref{fig-2}(b). These observations prove that the numerical simulations are in good agreement with the theory in the linear regime. 

Figure \ref{fig-2}(b) refers to {\it harmonic} and {\it subharmonic} tongues,
meaning responses to a forcing function with fundamental frequency
$\omega$ (period $T$) which have fundamental frequency $\omega$
(period $T$) and $\omega/2$ (period $2T$). Multiples of the
fundamental frequency are also present, but for simplicity, we will
refer to the fundamental frequency as the frequency.  For the
single-frequency Faraday instability, both responses exist, but the
most usual and prominent response is subharmonic.  For example, 
the simulations by \textcite{perinet2009numerical} of a hexagonal pattern show a
subharmonic response.  A harmonic response is sometimes possible;
see, e.g., \cite{kumar1996linear,muller1997analytic}, but it is exceptional.

The two-frequency forcing \eqref{eq:accel} for figure \ref{fig-2}(b) is a superposition of $\cos(4\omega t)$, with
weighting coefficient $\cos\chi$, and $\cos(5\omega t)$, with coefficient $\sin\chi$.
The usual subharmonic response to forcing frequency $4\omega$ would
be $2\omega$, while the usual subharmonic response to forcing $5\omega$ would
be $5\omega/2$. Two-frequency forcing yields a mixture of both responses \cite{besson1996two},
both of which are seen in Figure \ref{fig-2} for mixing angle $\chi=45^\circ$.
The large tongue shown in red is associated with forcing frequency $4\omega$
and has response $2\omega$, and the neighburing large tongue in black
is associated with forcing frequency $5\omega$ and has response frequency $5\omega/2$. 
Although the response frequency $2\omega$ is 
subharmonic in the context of forcing by single-frequency $4\omega$, it is called \textit{harmonic}
in the context of the two-frequency forcing function \eqref{eq:accel} because it is an
integer multiple of the overall frequency $\omega$.

Our main investigation, like that of
\textcite{arbell2002pattern}, has frequencies $(k,l) = (2,3)$ and mixing angle $\chi=30^\circ$. For this value of $\chi$, the $2\omega$-frequency
forcing, and hence the $\omega$ response, is dominant. Given the context of our two-frequency forcing, this response will be called harmonic.
We also see temporally subharmonic responses; indeed, the overall temporal dependence of the SSS-I pattern is subharmonic.

This nomenclature also applies to spatial dependence.
Given a wavenumber $k$, nonlinear interactions will necessarily create multiples of $k$ called superharmonics or merely harmonics,
merely by the elementary calculation $e^{ikx} e^{ikx}=e^{2ikx}$. 
When considering wavevectors 
%LSTAug29 UK orientated vs US oriented
%orientated 
oriented
in different directions, $2k$ must be replaced by vector sums $\bf{k}_1 +\bf{k}_2$. 
In the superlattice patterns that we investigate, modes with wavenumber $k/2$, which are {\it spatially subharmonic} to $k$, are also observed.
The mechanisms by which modes with linearly stable wavenumbers are generated and maintained are the topic of the many theoretical works cited in the introduction.

\begin{table*}[t]
  \def~{\hphantom{0}}
  \caption{\label{tab:grid}%
    Grid independence test in the linear regime: the method is
    repeated for two wavenumbers and grid convergence for the
    threshold amplitude (Comp.) is tested against the theoretical
    value (Theor.). The error is evaluated as
    $|F(\text{Theor.})-F(\text{Comp.})|/F(\text{Theor.})\times 100\,\%$.}
  \begin{ruledtabular}
    \begin{tabular}{lccccc}
      $k~(\mathrm{cm^{-1}})$ &
      $\lambda~(\mathrm{cm})$ &
      No.\ grid points $(\lambda/\Delta x)$ &
      $F$ (Theor.) &
      $F$ (Comp.) &
      Error (\%) \\
      \hline
      $10.258$ & $0.6125$ & $128$ & $3.377$ & $3.360$ & $0.503$ \\
      ~        & ~       & $~64$ & ~       & $3.357$ & $0.592$ \\
      ~        & ~       & $~32$ & ~       & $3.352$ & $0.740$ \\
      ~        & ~       & $~16$ & ~       & $3.177$ & $5.922$ \\
      \hline
      $8.096$ & $0.7830$ & $128$ & $2.118$ & $2.108$ & $0.472$ \\
      ~        & ~       & $~64$ & ~       & $2.102$ & $0.756$ \\
      ~        & ~       & $~32$ & ~       & $2.097$ & $0.992$ \\
      ~        & ~       & $~16$ & ~       & $2.046$ & $3.418$ \\
    \end{tabular}
  \end{ruledtabular}
\end{table*}

  \begin{figure*}
      \centering
      \includegraphics[width=0.9\linewidth]{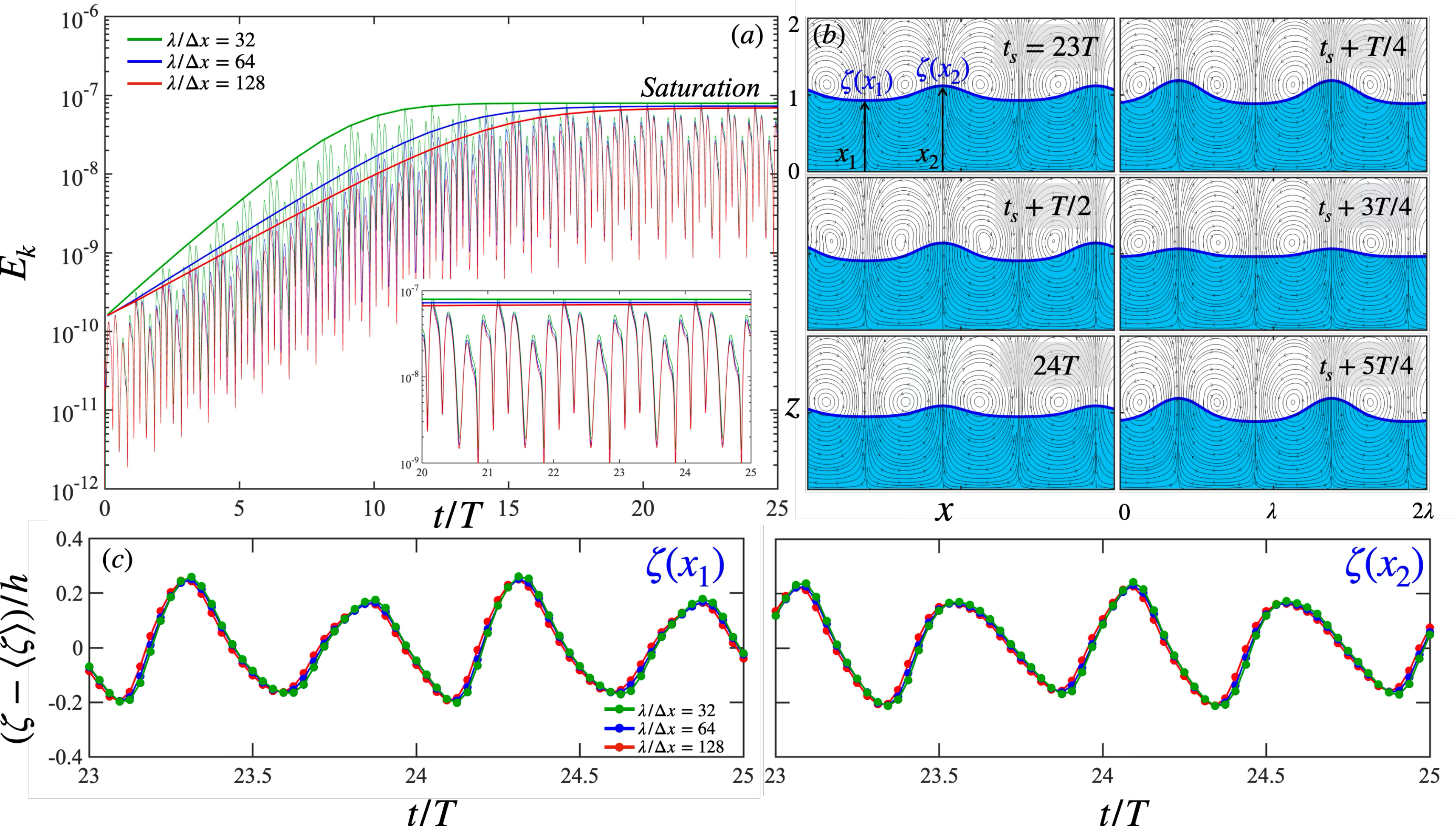}
      \caption{Grid independence test in the nonlinear regime: (a) Kinetic energy envelope for $F = 2.32$ for $32$, $64$, and $128$ grid points per wavelength. The inset is shown to magnify the evelope and the kinetic energy profile for $t \in [20, 25]$. Temporal evolution of the interface is shown in (b) for $\lambda/\Delta x = 64$, where the blue and white shades represent the liquid and gas phases and is separated by the interface, highlighted as thick deep blue line. The domain is overlaid with velocity streamlines. The temporal evolution of the interface position, $\zeta$ at position $x_1$ and $x_2$ are shown in (c) for $\lambda/\Delta x = 32, 64,$ and $128$. }
      \label{fig-3}
  \end{figure*}
  \setlength{\tabcolsep}{1pt}

A grid independence test is also carried out for the critical points of the harmonic and subharmonic tongues, as summarised in table \ref{tab:grid}. The threshold acceleration converges as the number of grid points per wavelength increases. For $32$ grid points, the error drops to below $1\%$ and the solution converges as we increase the resolution from $64$ to $128$ with an error of approximately $0.5\%$. This shows that at least $32$ grid points per wavelength are necessary to represent the two-frequency Faraday waves in the linear regime to 1\% accuracy. 

The next test for grid convergence is carried out in the nonlinear regime. We increase the acceleration above the threshold amplitude. Three mesh sizes are considered, $\lambda/\Delta x = 32, 64, $ and $128$, for which we set $\lambda = 0.783 ~\rm{cm}$ and $F = 1.1 \times 2.118$. The kinetic energy profiles and their envelope are shown in figure \ref{fig-3}(a). The linear growth of the kinetic energy eventually saturates due to nonlinear effects.
A significant difference is observed during the linear growth regime. However, as time progresses, 
saturation of the kinetic energy is observed in all 
three cases. The finer the grid resolution, the more time is taken to reach saturation. Once the envelope reaches the saturated state, an insignificant difference is observed in increasing the resolution, as shown in the inset of figure \ref{fig-3}(a). 

Snapshots of the $x$-$z$ slice at $y = \lambda/4$ for $t \in [23T, 24T+T/4]$ for $\lambda/\Delta x = 64$ are shown in figure \ref{fig-3}(b), where the velocity streamlines are overlaid on the slices. Qualitatively, we confirm that the standing interfacial waves are harmonic in nature (compare the snapshots at $t = 23T$ and $24T$). 
The interfacial height at one of the troughs ($x_1$) and crests ($x_2$) at $t = 23T$ is tracked for two time periods. This is shown in figure \ref{fig-3}(c). We observe that the interfacial height, $\zeta$, at $x_1$ and $x_2$, repeats after one time period, confirming that the waves are harmonic in nature. Not only the kinetic energy profile, but also the interfacial heights, $\zeta(x_1)$ and $\zeta(x_2)$, show negligible differences in increasing the grid resolution from $32$ to $128$. Therefore, a 
resolution of $32$ grid cells per wavelength seems to be adequate at these parameter values for the nonlinear regime as well.

The final validation test uses parameters that match those of the 
experiments \cite{arbell2002pattern}.
The density and dynamic viscosity of the silicone oil are set to $950 ~\rm{kg/m^3}$ and $0.0219 ~ \rm{Pa.s}$. The surface tension is set to $21.5 \times 10^{-3}~ \rm{N/m}$. The fluid is filled to a height of $h = 0.155~ \rm{cm}$, and the height of the computational domain is $H = 1 ~ \rm{cm}$. The parameters for the imposed oscillation \eqref{eq:accel} are the fundamental \rv{driving} frequency $\omega = 2\pi(20) \rm{rad ~s^{-1}}$, 
the two oscillation frequencies $2\omega$ and $3\omega$, the mixing angle $\chi = 30^\circ$, and the phase angle $\phi = 0^\circ$. 
%LSTJun24 added
At mixing angle $\chi=30^\circ$, the $4\omega$-frequency forcing, and hence the $2\omega$ response, is dominant.
For these parameters, the critical wavelength is $\lambda_c=0.785~\rm{cm}$, 
and we set the lateral size of the domain to 
%LSTJul10 Why does y direction look longer than x direction???
$4\lambda_c/\sqrt{3} \times 4\lambda_c$, as illustrated in figure \ref{fig-1}.
The acceleration amplitude is set to $F = 3.46$, which we will denote by $F_h$. At this amplitude, 
\textcite{arbell2002pattern} determined experimentally 
that the surface organises into hexagonal patterns. 

The interface is initialized with stripes as $\zeta = 0.05h \,(\cos k_cy)$, where $k_c = 2\pi / \lambda_c$.
After the initial saturation of the stripes, the final state is the hexagonal interface shown in figure \ref{fig-1}(a);  the temporal evolution of the kinetic energy is shown in figure \ref{fig-1}(b). 
Figure \ref{fig-1}(b) also shows that the hexagons persist for at least $100\, T$, during which we observed no secondary instability. 
This contrasts with \cite{perinet2012alternating}, for which 
hexagonal patterns lasted only 10 subharmonic periods before transitioning to more complicated dynamics.
Initialising
 instead
with a planar surface perturbed only by numerical error also led to the formation of a stable hexagonal pattern. 
This leads us to believe that the formation of these hexagonal patterns is independent of the initial state of the system.

\subsection{Fourier space analysis}
\label{sec:fourier}

\begin{figure*}
    \centering
    \includegraphics[width=0.85\linewidth]{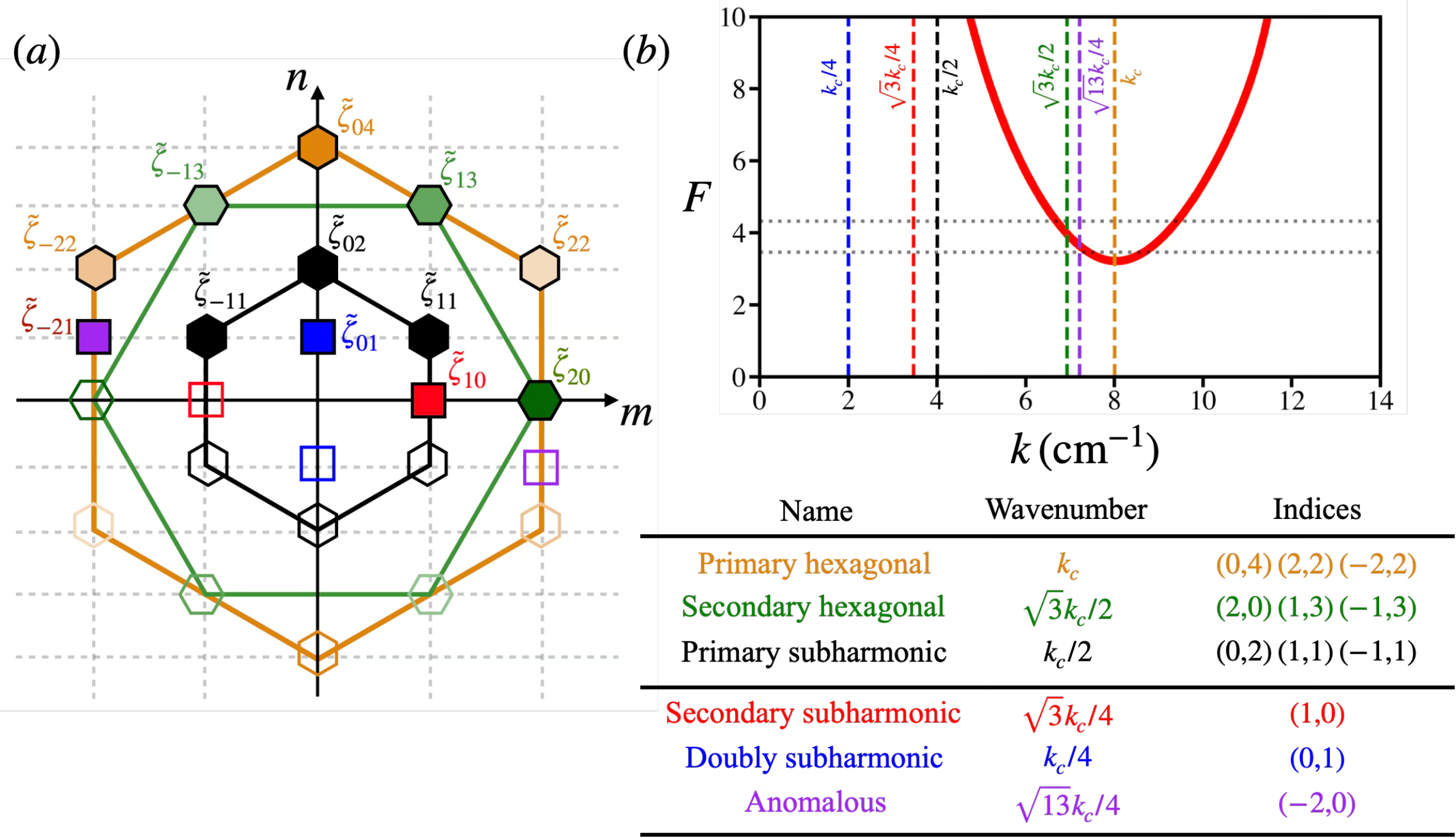}
    \caption{(a) The spatial Fourier space grid with relevant modes. The color coding of the markers will be used thoughout the manuscript. 
    In addition to the {\it primary hexagonal} modes with critical wavenumber $k_c$ shown in orange, two other hexagons with smaller wavenumbers (larger wavelengths) are present, the {\it secondary hexagonal} with $k=\sqrt{3} k_c/2$ shown in green, and the {\it primary subharmonic} ones with $k=k_c/2$ shown in black.
    Three more {\it non-hexagonal} modes, shown in red, blue, and purple, play crucial roles in the patterns. These modes have two indices of opposite parity and so cannot be part of a hexagonal triad on our rectangular Fourier grid, as demonstrated in Appendix \ref{ss-A1}. (b) Linear instability tongue for the experimental configuration
    of \textcite{arbell2002pattern}.
    The temporally subharmonic tongue boundary is 
    the thick red curve, inside of which is the linearly unstable regime,
    i.e. parameter pairs $(k,a/g)$ such that the flat surface vibrated with amplitude $a$ is unstable to perturbations with wavenumber $k$. Other instability tongues exist, but with minima above $F=10$.
    Physical parameters are $\rho_l = 950~\rm{kg/m^3}$, $\mu_l = 0.0219~\rm{Pa.s}$, and $\sigma = 21.5\times 10^{-3}~\rm{N/m}$. Forcing parameters are $\omega/2\pi = 20~\rm{Hz}$, $(k,l) = (2,3)$, and $\chi = 30^\circ$. }
    \label{fig:SL-A}
\end{figure*}

To quantitatively analyse the pattern, we perform a spatial Fourier analysis of the two-dimensional interfacial elevation $\zeta(\mathbf x, t)$, defined by
\begin{subequations}
\begin{align}
    \zeta(\mathbf{x}, jT) &= \sum_{m,n} e^{i \mathbf{k}_{mn} \cdot \mathbf{x}} 
    \tilde{\zeta}_{m,n}(jT), \quad j=0, 1, \ldots \\
    \hat{\zeta}_{m,n} &= |\tilde{\zeta}_{mn}|
\end{align}
\label{eq:FT}
\end{subequations}
Because the interface $\zeta(x,y)$ is real, the complex amplitude $\tilde\zeta_{-m,-n}$ is the complex conjugate of $\zeta_{m,n}$
By a slight abuse of notation, in what follows we will use $\hat\zeta_{m,n}$ to designate both the name of the Fourier mode and also its real amplitude. We will retain $\tilde\zeta_{m,n}$ to refer to its complex amplitude. 

The decomposition is carried out over the rectangular computational domain, constraining the admissible wave vectors $\mathbf k_{mn}$ to lie on a grid:
\begin{equation}
\mathbf k_{m,n} = \left(\sqrt{3} m~ \mathbf e_x + n ~\mathbf e_y\right)k_c/4,  ~~\forall (m,n) \in \mathbb Z,
\end{equation}
where $k_c$ is the critical wavenumber. Thus, 
\begin{equation}
|k_{m,n}| = \left(3 m^2 +n^2 \right)^{1/2}k_c/4.
\end{equation}
Figure \ref{fig:SL-A}(a)
shows this grid, as well as the non-negligible Fourier modes governing the dynamic behaviour we have observed in our simulations. 

Three sets of modes form hexagons. These modes are shown as hexagonal markers connected by solid lines.
The \emph{primary hexagonal modes} $\hat \zeta_{-2,2}$, $\hat \zeta_{0,4}$, $\hat \zeta_{2,2}$ have wavenumber $k_c$ and are shown in orange. The \emph{secondary hexagonal modes}, $\hat \zeta_{-1,3}$, $\hat \zeta_{1,3}$, $\hat \zeta_{2,0}$ have a smaller wavenumber $\sqrt {3}k_c/2$ and are shown in green. Figure \ref{fig:SL-A}(b) shows the results of linear stability analysis of the flat interface for the parameters of \cite{arbell2002pattern}.
In Fig. \ref{fig:SL-A}(b),  $\chi=30^\circ$ so, unlike in Fig.\ \ref{fig-2} for $\chi=45^\circ$, we are far from the codimension-two point and so the subharmonic tongue (and all other tongues) is above $F=10$ and so outside of the range of the figure. 
Vertical dashed lines are drawn at the values of $k$ for which markers are  drawn in \ref{fig:SL-A}(a), with the same color coding.
The primary hexagonal modes are unstable for $F \geq F_h$, while the secondary hexagonal modes become unstable for  $F\geq 1.25\,F_h$.
The pattern also contains a set of hexagonal modes which have wavenumber $k_c/2$, the {\it primary subharmonic modes} $\zeta_{0,2}$, $\hat \zeta_{1,1}$ and $\hat\zeta_{-1,1}$. Shown in black in figure \ref{fig:SL-A}(a), these are linearly stable modes of the flat surface.
%and so must arise via other means, such as nonlinear interactions or secondary instability of the hexagons.

Three other modes appear with appreciable amplitude in our simulations, all shown as square markers. 
Mode $\hat\zeta_{1,0}$, with wavenumber $\sqrt{3}\,k_c/4$ and shown in red, is subharmonic to the secondary hexagonal modes; we will call it a {\it secondary subharmonic} mode. 
Mode $\hat\zeta_{0,1}$ with wavenumber $k_c/4$ is shown in blue. It is a subharmonic of the primary subharmonic modes and so we have called it {\it doubly subharmonic}. 
Finally, $\hat \zeta_{-2,1}$, of wavenumber $\sqrt{13}\,k_c/4$, is shown in purple and will be called {\it anomalous}.  All three of these modes have indices of {\it mixed parity} so that their sum is odd, i.e. $-2+1=-1$ and $1+0=1$. Appendix \ref{ss-A1} shows that their images under rotation by $\pi/3$ do not have integer indices and so cannot be represented on the rectangular grid. Such points therefore cannot belong to a hexagon on the grid and we call such modes {\it non-hexagonal}. Of the three, $\hat\zeta_{1,0}$ and $\hat\zeta_{0,1}$ fall outside the instability tongue of figure \ref{fig:SL-A}(b), while $\hat\zeta_{-2,1}$ falls inside. 

 These three non-hexagonal modes all play important roles, as will be shown in \S \ref{ss_DNA}. The secondary subharmonic mode $\hat\zeta_{1,0}$ is a crucial ingredient of our SSS-I pattern; see Fig.
\ref{fig:Arbell}(b).
The doubly subharmonic mode $\hat\zeta_{0,1}$ will be seen in \S \ref{sec:dyn} to play an important role in its long-time dynamics. The anomalous mode $\hat\zeta_{-2,1}$ plays a prominent, though transient, role in the transition to SSS-I in the simulation of \S \ref{ss-short}. This mode also appears in a modulated regime found for lower vibrational amplitude, described in Appendix \ref{sec:lowerF}.

\section{Superlattice state}
\label{ss_DNA}
\begin{figure*}
    \centering    \includegraphics[width=1\linewidth]{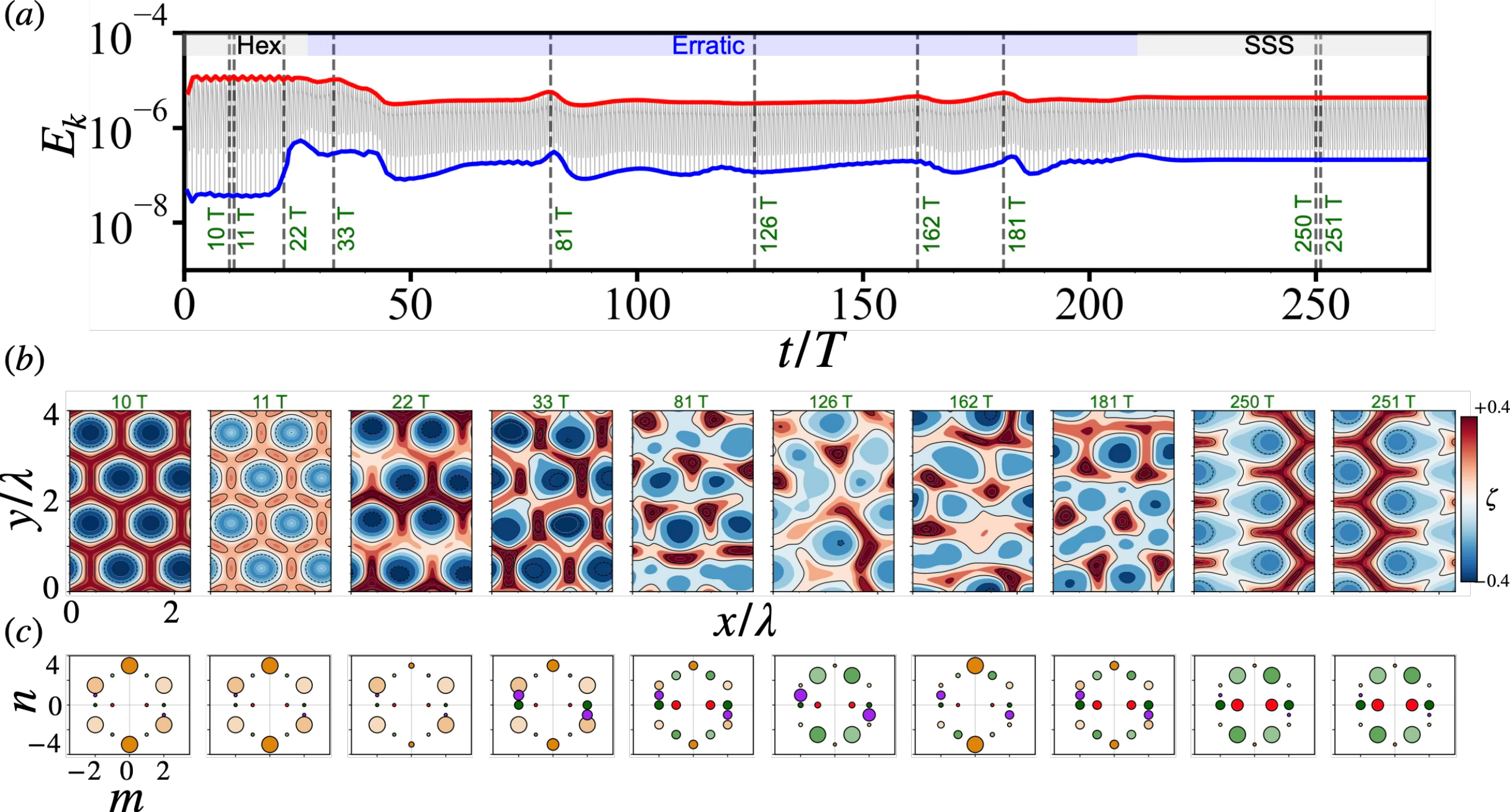}
    \caption{(a) Envelope of maximum and minimum kinetic energy time series. (b) Interface at instants 
    indicated by  dotted lines in $E_k$ profile. (c) 
    Spatial spectra at the same instants, where the color code is that of figure \ref{fig:SL-A}, and the size of the markers indicates the mode amplitude. The initial pattern at $10\,T$ consists of primary hexagonal modes (orange). The subharmonic response in the hexagons can be seen by comparing $10\,T$ with $11\,T$. By $126\,T$,  secondary hexagonal modes (green) have appeared, and by $250\,T$, the secondary subharmonic mode (red) responsible for the ``backbone'' is present. The temporally subharmonic (period $T$) response in the final SSS-I pattern can be seen as the flipping from right to left of the backbone between $250\,T$ and $251\,T$. 
    }
    \label{fig:SL-B}
\end{figure*}

We turn our discussion to the superlattice wave patterns, which arise out of an instability of a hexagonal pattern. 
We set the forcing amplitude to $F = 1.25\, F_h$ and performed simulations starting from two distinct initial conditions. 
For the first run,  we used the nonlinear hexagonal state at $F=F_h$ as the initial condition. For the second run, we applied random noise to a small-amplitude hexagonal perturbation. In both cases, we obtained the same superlattice pattern but by different routes. We present the results of these two simulations below.
We also ran simulations at lower values of the forcing amplitude $F=1.05\,F_h$ and $F=1.10\, F_h$, which did not lead to the superlattice pattern. We defer describing these simulations at lower $F$ to the Appendix \ref{sec:lowerF}. 

\subsection{Simulation starting from  hexagonal equilibrium at \texorpdfstring{$F = F_h$}{}}
\label{ss-short}

We used as the initial condition the hexagonal state obtained in \S~\ref{ss_validation} with $F = F_h$ at $t = 350\,T$. The velocity, pressure, and Lagrangian surface fields were saved, the clock was reset to $t = 0$, and the acceleration amplitude was increased to $F = 1.25\,F_h$. 
The kinetic energy envelope is shown in figure \ref{fig:SL-B}(a), and the evolution of the interface heights at different driving periods is shown in figure \ref{fig:SL-B}(b).
The sudden increase in $F$ leads abruptly to hexagons of higher kinetic energy.  
The first feature observed is an oscillation of period $2T$ between two very close values of $E_k$ 
in figure \ref{fig:SL-B}(a). Figure \ref{fig:SL-B}(b) shows the patterns at $t = 10T$ and $11T$. Both are hexagonal, but they differ slightly. The heights, indicated by the color code, differ between the two instants and at $t = 11T$, a small bump appears at the bottom of the trough and protrusions form on the six edges of the crown surrounding the trough.
\begin{figure*}
    \centering
    \includegraphics[width=0.7\linewidth]{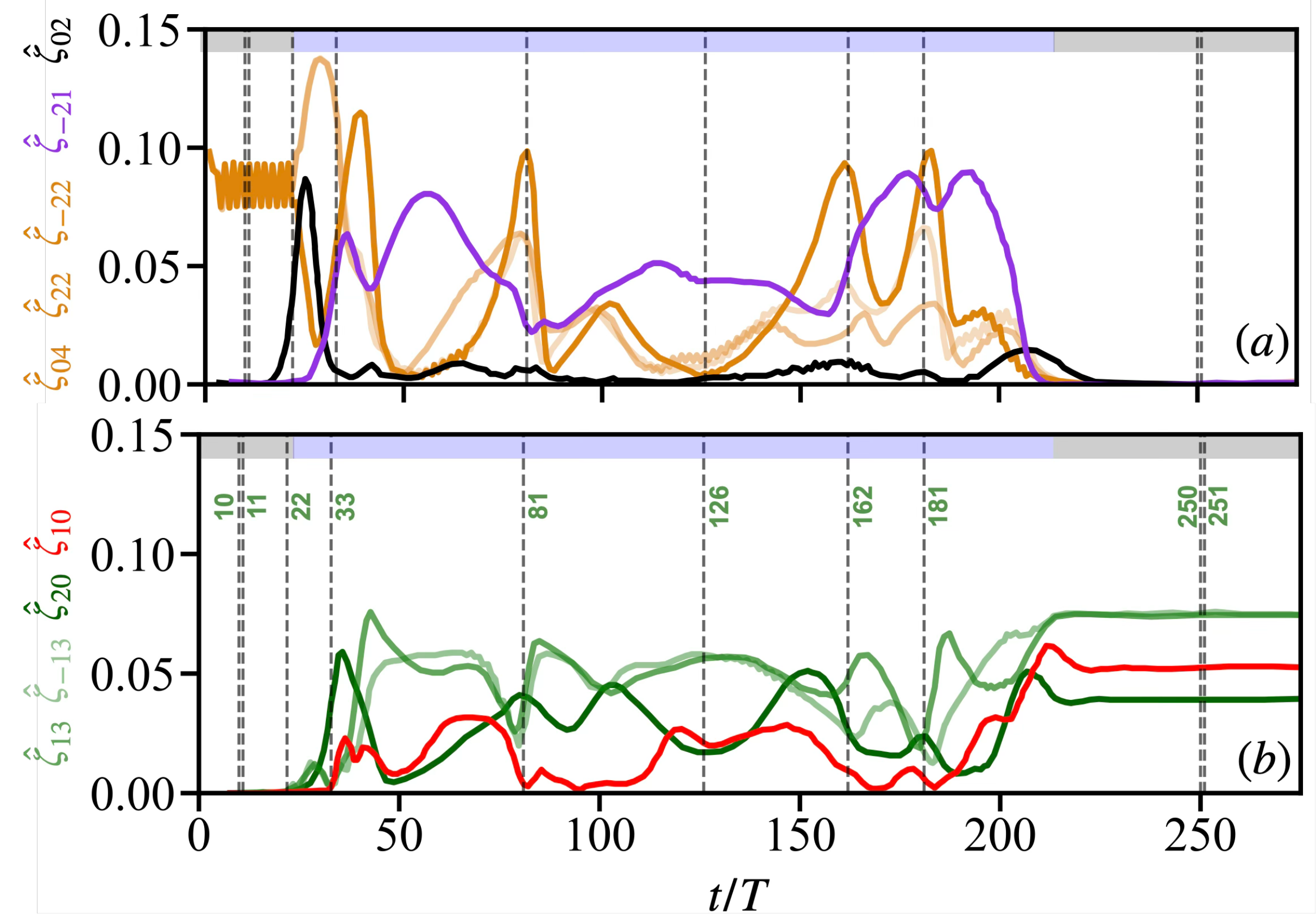}
    \caption{
    Temporal evolution of the spatial Fourier coefficients. Curves are color-coded according to the Fourier space grid in figure \ref{fig:SL-A}(a).
    (a) Primary hexagonal (orange) modes comprising initial hexagonal pattern together with primary subharmonic  (black, $\hat\zeta_{0,2}$) and anomalous modes (purple, $\hat\zeta_{-2,1}$). (b) Secondary hexagonal modes (green) together with secondary  subharmonic (red, $\hat\zeta_{1,0}$) which comprise the SSS-I pattern.}
    \label{fig:SL-C}
\end{figure*}

Figures \ref{fig:SL-B}(c) and \ref{fig:SL-C} show the time evolution of the amplitudes of the significant Fourier modes. The anomalous mode $\hat\zeta_{-2,1}$ rises (figure \ref{fig:SL-C}(c)) and persists during the erratic phase alongside the broader mode competition. Figure \ref{fig:SL-B}(c) represents the mode amplitudes via the size of points on the Fourier grid at various instants, while Figure \ref{fig:SL-C} shows their time evolution as curves grouped by type, with subfigure (a) for the primary hexagonal modes, (b) for the secondary hexagonal modes and the secondary subharmonic mode $\hat\zeta_{1,0}$, and (c) for the primary subharmonic hexagonal modes and the anomalous mode $\hat\zeta_{-2,1}$.

The short-lived temporally subharmonic hexagonal state can be seen at the beginning of figure \ref{fig:SL-C}(a), where the three primary hexagonal modes all oscillate together with period $2T$, with a peak-to-peak amplitude of approximately $20\%$ of their temporal mean.  
Since all modes surveyed other than those of the primary hexagon are negligible at this time, the changes we see in figure \ref{fig:SL-B}(b) between $10T$ and $11T$ must be due to higher spatial harmonics that oscillate subharmonically.

The temporally subharmonic hexagons last until $t\approx 22 T$. 
After this, the kinetic energy envelope, the patterns, and the Fourier spectra all change dramatically, signalling the start of the erratic dynamics. 
The hexagonal symmetry is lost. In figure \ref{fig:SL-C}(a), the amplitude of $\hat\zeta_{0,4}$ decreases far below that of the other two primary modes, and its subharmonic $\hat\zeta_{0,2}$ in figure \ref{fig:SL-C}(c) rises, along with the other two primary subharmonic modes.
This is followed by an increase in the secondary hexagonal and the non-hexagonal modes in figures \ref{fig:SL-C}(b) and (c). The anomalous mode $\hat\zeta_{-2,1}$ rises (figure \ref{fig:SL-C}(c)) and persists during the erratic phase alongside the broader mode competition.  Recall that the secondary hexagonal modes and $\hat\zeta_{-2,1}$ are all linearly unstable modes of the flat surface for $F=1.25\,f_h$; see figure \ref{fig:SL-A}.
%Period $2T$ oscillations of the primary modes are superimposed on the smoother dynamics throughout the transient; however, given their small amplitudes, they barely alter the shape of the interface. 
%
Between $t = 81\, T$ and $181\,T$, the Fourier spectra show the simultaneous presence of the primary, secondary, and primary subharmonic hexagonal modes, as well as two non-hexagonal modes, giving a total of $11$ non-negligible modes, plus their complex conjugates. Eventually, at $t \geq 220\,T$, the simulation converges towards a stationary state in which two of the secondary hexagonal modes $\hat\zeta_{\pm 1, 3}$ dominate $\hat\zeta_{2,0}$, with the secondary harmonic mode $\hat\zeta_{1,0}$ between them. All other modes are of negligible amplitude. 
%LSTAug6 can say some other time perhaps
%The linear stability analysis of the flat surface shown in figure \ref{fig:SL-A}(b) shows that the secondary hexagonal modes are excited, but $\hat\zeta_{1,0}$ is not. However, the hexagons at $F=1.25F_h$ are of sufficiently high amplitude that such an analysis is irrelevant.

This is the spectral composition of the superlattice SSS-I pattern presented by \textcite{arbell2002pattern} and reproduced in figure \ref{fig:Arbell}, with a crucial modification. The modes present in the SSS-I of \cite{arbell2002pattern} are the three {\it primary} hexagonal modes ($k=k_c$) and one {\it primary} subharmonic mode ($k=k_c/2$), whereas those in figure \ref{fig:SL-B}(c) are the {\it secondary} hexagonal ($k=\sqrt{3}k_c/2$) and {\it secondary} subharmonic ($k=\sqrt{3}k_c/4$) modes. We do not know the reason for this difference.

The interface goes through a wide variety of patterns, as shown in the snapshots of figure \ref{fig:SL-B}(b).
%LSTAug7 
%The presence of $\hat\zeta_{2,0}$ and $\hat\zeta_{1,0}$ produces a long-wavelength modulation ultimately giving rise to the 
The final 
superlattice pattern at $250\,T$ 
features a brown backbone or skeleton centered on $x/\lambda \approx 2$ with period $4/3$ in $y/\lambda$, reminiscent of strands of DNA.

\begin{figure*}
    \centering    \includegraphics[width=\linewidth]{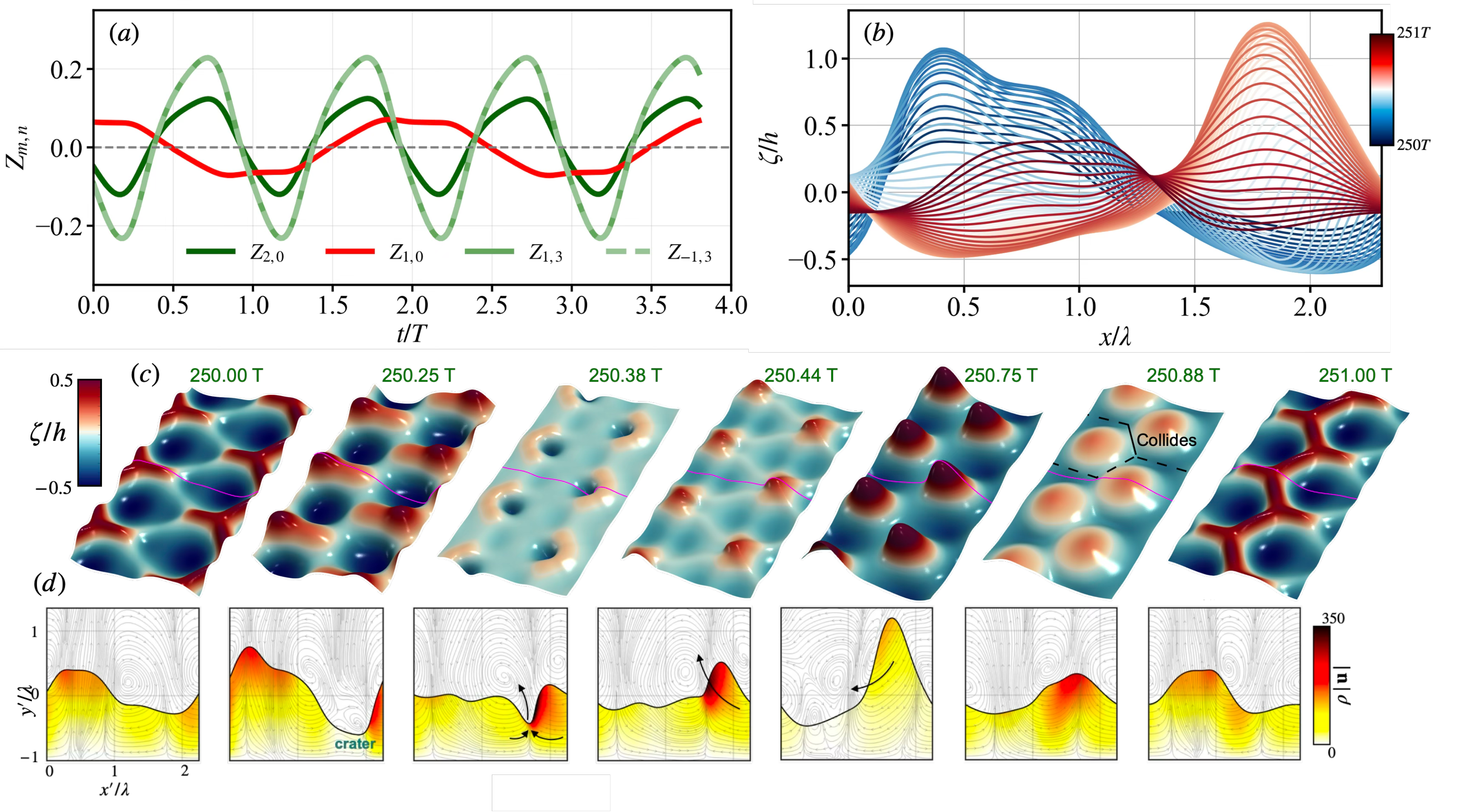}
    \caption{Temporal behavior of SSS-I pattern. (a) Secondary hexagonal and subharmonic modes comprising the SSS-I pattern over $4T$. Here $Z_{m,n} \equiv \hat \zeta_{m,n}~\mathrm {sgn}(\mathrm{Re}(\tilde \zeta_{m,n})$). The secondary hexagonal modes have period $T$ (temporally harmonic), while the secondary subharmonic mode has period $2T$ (temporally subharmonic). (b,c,d) Temporal behavior over $[250\,T, 251\, T]$. (b) Height of interface along slice $S$ indicated in (c), interface. (d) Two-dimensional projection along $S$.  The difference between the interface at $t=250$ and $t=251$ is a manifestation of the fact that $Z_{1,0}(251)=-Z_{1,0}(250)$.
     }
    \label{fig-10}
\end{figure*}

We examine the detailed spatiotemporal dynamics of our superlattice pattern in figure \ref{fig-10}. 
We present the temporal evolution of its four Fourier modes over four time periods in figure \ref{fig-10}(a). 
In past representations of the Fourier components, we were concerned with the long-term evolution of the pattern and hence plotted the amplitudes $\hat\zeta_{m,n}$ at integer multiples of $T$. Now, we wish to show the dynamics of the pattern within one or a few periods, and so we plot instead $Z_{m,n} \equiv \hat \zeta_{m,n}~\mathrm {sgn}(\mathrm{Re}(\tilde \zeta_{m,n})$. 
It can be seen that, while the hexagonal modes have a period of $T$, the spatially subharmonic mode $\hat\zeta_{1,0}$ is also temporally subharmonic, i.e.\ $2\,T$-periodic, 
so the overall dynamics is also $2\,T$-periodic. This is also the case in \cite{arbell2002pattern} and is indeed the reason for which they chose the name of Subharmonic Superlattice State.
%As in \cite{arbell2002pattern}, the mode $\hat\zeta_{1,0}$ oscillates subharmonically in time whereas the secondary hexagonal modes behave harmonically. 
%LSTJul9 Don't forget to change last label from 250 to 251. And the first panel, 250, looks a bit strange -- smaller, different angle. 

The other panels of figure \ref{fig-10} display visualizations of the interface over $[250\,T,251\,T]$, with figure \ref{fig-10}(c) showing the three-dimensional interface and Figures \ref{fig-10}(b,d) showing projections in the vertical plane section $S$. 
The interfaces at $t = 250\,T$ and $251\,T$ have the same shape, but are images of one another by a half-wavelength translation in $x$: the brown ``backbone'' is at the edges in $x$ at $250\,T$ and at mid-$x$ at $251\,T$. This is due to the contribution by $\hat\zeta_{1,0}$ with a temporal period of $2\,T$.

%Our simulations have provided results that are in partial agreement with the experimental findings of \cite{arbell2002pattern}. The spatio-temporal features of both states are globally the same, but the scale of the simulated patterns differs from the experimental ones by a factor of $\sqrt{3}/2$. 
%We do not know the reason for this difference.

Our simulations have thus produced a 
superlattice pattern that has the same shape and a Fourier spectrum similar to that of the SSS-I state of  \textcite{arbell2002pattern}, but with some noticeable differences. Their SSS-I pattern is composed of the three primary modes ($k=k_c$) and one primary subharmonic mode ($k=k_c/2$). In contrast, our SSS-I-like pattern contains the three secondary modes ($k=\sqrt{3}k_c/2$) together with $\hat{\zeta}_{1,0}$, which is a spatial subharmonic of the secondary modes ($k=\sqrt{3}k_c/4$).
The temporal oscillation of our hexagonal and spatially subharmonic modes is harmonic and subharmonic, respectively, like that of \cite{arbell2002pattern}.

\subsection{Simulation starting from a perturbed flat interface - long-time behaviour of superlattice wave-patterns}
\label{ss_LONG}
\begin{figure*}
    \centering
    \includegraphics[width=1.0\linewidth]{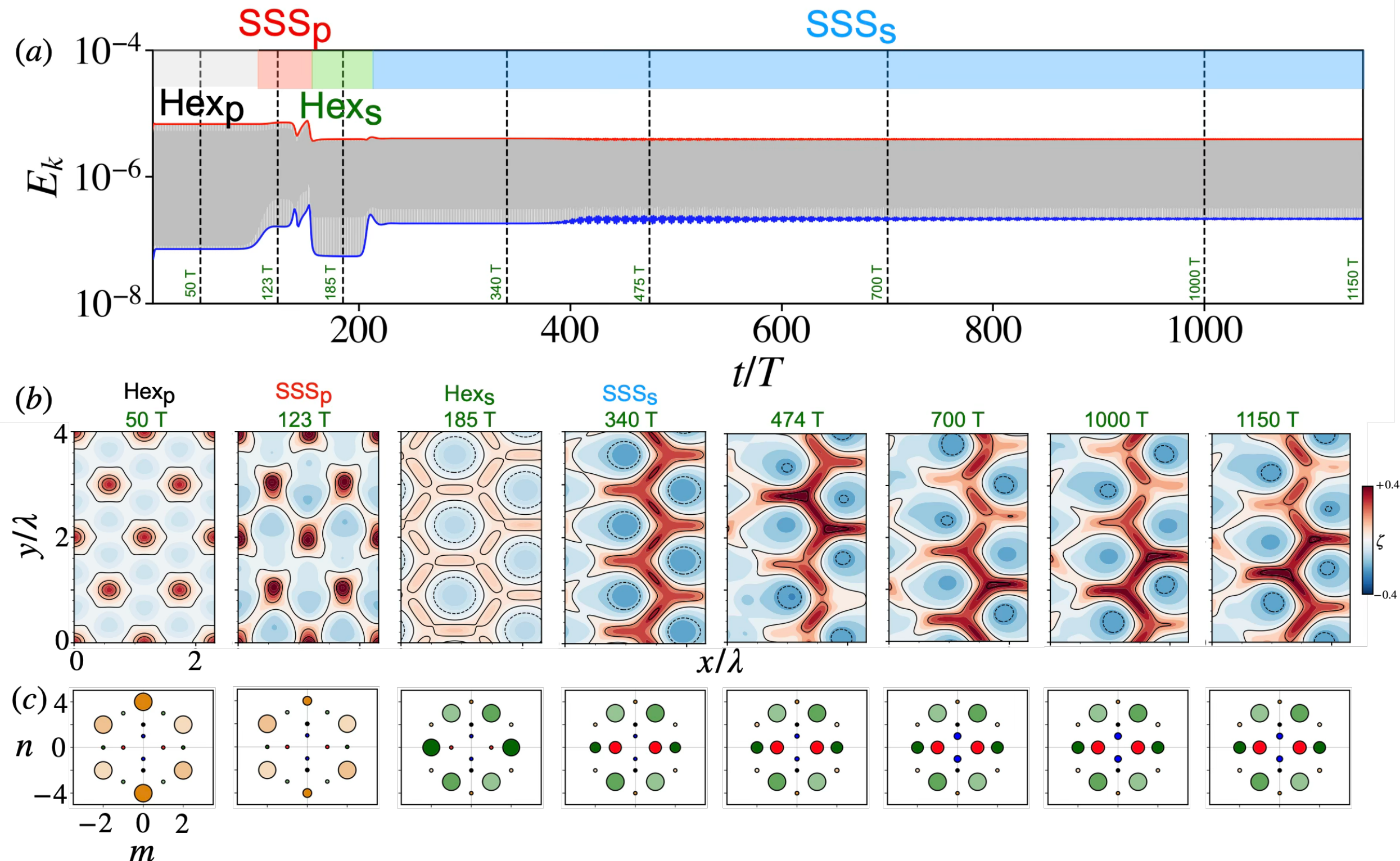}
    \caption{(a) 
    Envelope of maximum and minimum kinetic energy time series over $t \in [0, 1150\,T]$ for $F=1.25\,F_h$, starting from initial condition \eqref{eq:init_cond}. Four sequential regimes are identified and shaded: the primary hexagonal state 
    $\mathrm{Hex_p}$, the  primary Subharmonic Superlattice State $\mathrm{SSS_p}$, the secondary hexagonal state $\mathrm{Hex_s}$, and the secondary Subharmonic Superlattice State $\mathrm{SSS_s}$. Vertical dashed lines indicate time instants shown in (b) and (c). (b) Contour plots of the interface height $\zeta$ at times indicated above. The first four snapshots are from each of the four regimes, $\mathrm{Hex_p}$, $\mathrm{SSS_p}$, $\mathrm{Hex_s}$, and $\mathrm{SSS_s}$. The four later snapshots show the dynamic SSS state, illustrating a vertical drift of the horizontal arms emanating from the main branch caused by $\hat\zeta_{\pm1,3}$ and the modulation in intensity along the backbone driven by $\hat\zeta_{0,1}$.
    (c) Spatial spectra at the same instants, where the color code is that of figure \ref{fig:SL-A}, and the size of the markers indicates the mode amplitude. Hex$_p$ contains the primary hexagonal modes (orange), one of which decreases in going to SSS$_p$. These are succeeded by the secondary hexagonal modes (green) forming  Hex$_s$, one of which diminishes while the secondary subharmonic (red) appears, yielding SSS$_s$. The remaining spectra correspond to the dynamic SSS and change very little throughout the evolution. 
    %LSTAug26 -- check??
    %because the pattern dynamics are due primarily to the phases of the modes.
    }
    \label{fig:sss_1}
\end{figure*}

In \S~\ref{ss-short}, we obtained a superlattice wave-pattern and compared it with the SSS-I state obtained by 
\cite{arbell2002pattern}. In this section, we carry out a longer simulation with the same parameters but have chosen a different initial condition: 
a low-amplitude hexagonal state perturbed with random noise
$\eta(x,y)$:
\begin{equation}
    \frac{\zeta(x,y)}{h} =
0.05\sum_{j=0}^{2} \cos\left(k_c~x \cos\frac{2\pi j}{3} + k_c~y \sin\frac{2\pi j}{3}\right) + \eta(x,y).
    \label{eq:init_cond}
\end{equation}
We will see that the simulation eventually leads to the same SSS-I state as in section \S~\ref{ss-short} but with a different pathway. %However, the path taken by the evolution is quite different. 

\begin{figure*}
    \centering
    \includegraphics[width=0.8\linewidth]{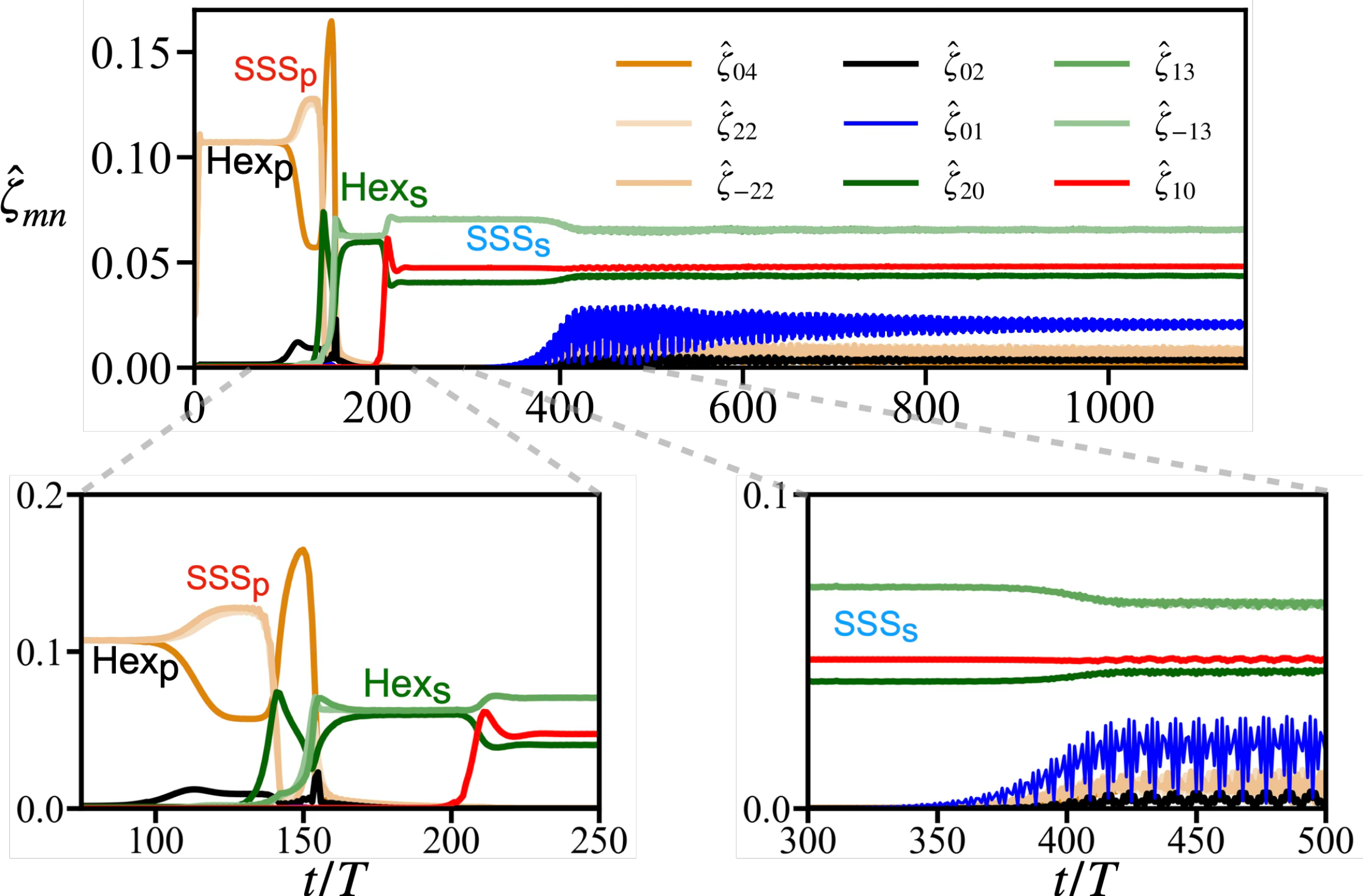}
    \caption{Temporal evolution of the Fourier mode amplitudes $\hat\zeta_{mn}$ for $F = 1.25F_h$. (a) Full interval $t \in [0, 1150T]$: primary modes ($\hat\zeta_{0,4}$, $\hat\zeta_{2,2}$, $\hat\zeta_{-2,2}$), secondary modes ($\hat\zeta_{1,3}$, $\hat\zeta_{-1,3}$, $\hat\zeta_{2,0}$), primary subharmonic mode $\hat\zeta_{0,2}$ and non-hexagonal modes $\hat\zeta_{1,0}$ and $\hat\zeta_{0,1}$. %(respective wavenumbers $\sqrt{3}k_c/4$ and $k_c/4$). 
    (b) Magnified view of $t \in [0, 205\,T]$ showing the $\mathrm{Hex_p} \to \mathrm{SSS_p} \to \mathrm{Hex_s}$ transition: the three primary modes are equal in $\mathrm{Hex_p}$, $\hat\zeta_{0,4}$ rises sharply to form $\mathrm{SSS_p}$, and the secondary modes grow to establish $\mathrm{Hex_s}$. (c) Magnified view of $t \in [250\,T, 600\,T]$ showing the onset at $t \approx 350\,T$ of the dynamic version of the $\mathrm{SSS_s}$ state: the dominant secondary modes and $\hat\zeta_{1,0}$ develop slight oscillations, while $\hat\zeta_{0,1}$ grows from zero  and develops sustained large-amplitude oscillations.}
    \label{fig:sss_2}
\end{figure*}

The envelopes of the kinetic energy and the amplitudes of the main Fourier modes are shown in figures \ref{fig:sss_1}(a) and \ref{fig:sss_2}, respectively. 
As in section \S~\ref{ss-short}, the kinetic energy seen in figure \ref{fig:sss_1}(a) initially rises and saturates at a hexagonal state consisting of primary modes, which we call $\mathrm{Hex_p}$. 
The state $\mathrm{Hex_p}$ differs from the initial hexagonal state seen in \S~\ref{ss-short} in several  ways. 
First, unlike the temporally subharmonic oscillations (period $2\,T$) observed in \S~\ref{ss-short}, the hexagons here oscillate harmonically (period $T$) and hence are seen as stationary in figures \ref{fig:sss_1} and \ref{fig:sss_2} over approximately $100\,T$. 
Second, the kinetic energy of $\mathrm{Hex_p}$ is lower than that of the corresponding state in section \S~\ref{ss-short}. 
This can be seen by comparing the initial portions of figures \ref{fig:sss_1}(a) and \ref{fig:SL-B}(a) or those of figures \ref{fig:SL-C}(a) and \ref{fig:sss_2}.

As was previously the case, 
the transition from $\mathrm{Hex_p}$ is initiated when the amplitudes of the three primary modes become unequal and the spatial subharmonic 
$\hat\zeta_{0,2}$ rises, here at $t\approx 100\,T$, as seen in figure \ref{fig:sss_2}.
At later times, the simulations behave quite differently.
While the evolution in \S~\ref{ss-short} proceeds via a long erratic phase, here it takes place through a number of quasi-stationary phases, like those manifested when passing through unstable steady states. 
It is for this reason that we have labelled and named the intermediate phases of this evolution.
We call the next state observed a (primary) superlattice state $\mathrm{SSS_p}$ because of its zig-zag appearance and the fact that it consists of the primary hexagonal modes together with a primary subharmonic $\hat\zeta_{0,2}$. 

Subsequently, a rapid and dramatic change occurs as these primary modes decay and are replaced by the secondary hexagonal modes, which grow to establish a near-hexagonal state $\mathrm{Hex_s}$. (Its slight departure from hexagonal symmetry can be seen by the fact that $\hat\zeta_{2,0}$ is slightly lower than the other secondary modes and the hexagonal crowns of \ref{fig:sss_1}(b) at $t = 185\,T$ are not perfectly regular.) At $t \approx 200\,T$, the secondary subharmonic mode $\hat\zeta_{1,0}$ rises, and by $t \approx 340\,T$, a superlattice has formed that contains the secondary modes together with a spatial subharmonic, which we now call $\mathrm{SSS_s}$. This is the same state as in \S~\ref{ss-short}, 
more specifically figure \ref{fig:SL-B}(b) at $t=250\,T$.

\begin{figure}
    \centering
%  Markers for \phi_01 for Nicolas. 
    %  t = 800-1000 T 
    \includegraphics[width=1.0\linewidth]{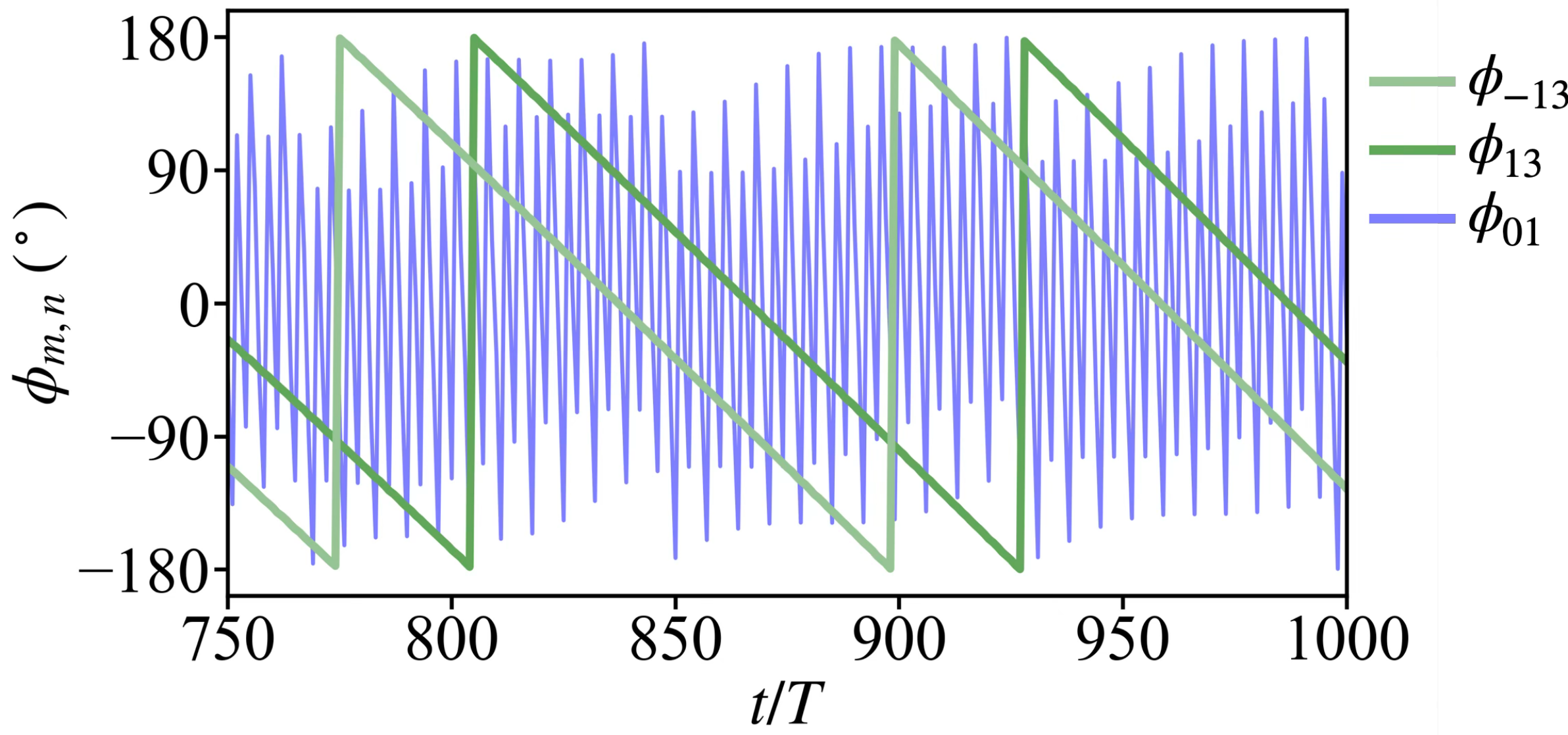}
    \caption{Temporal evolution of some spatial phases.
      $\phi_{0,1}$ oscillates rapidly, while $\phi_{1,3}$ and $\phi_{-1,3}$ decrease at the same slow and constant speed.}
    \label{fig:del_phi}
\end{figure}
\begin{figure*}
    \centering
    \includegraphics[width=1.0\linewidth]{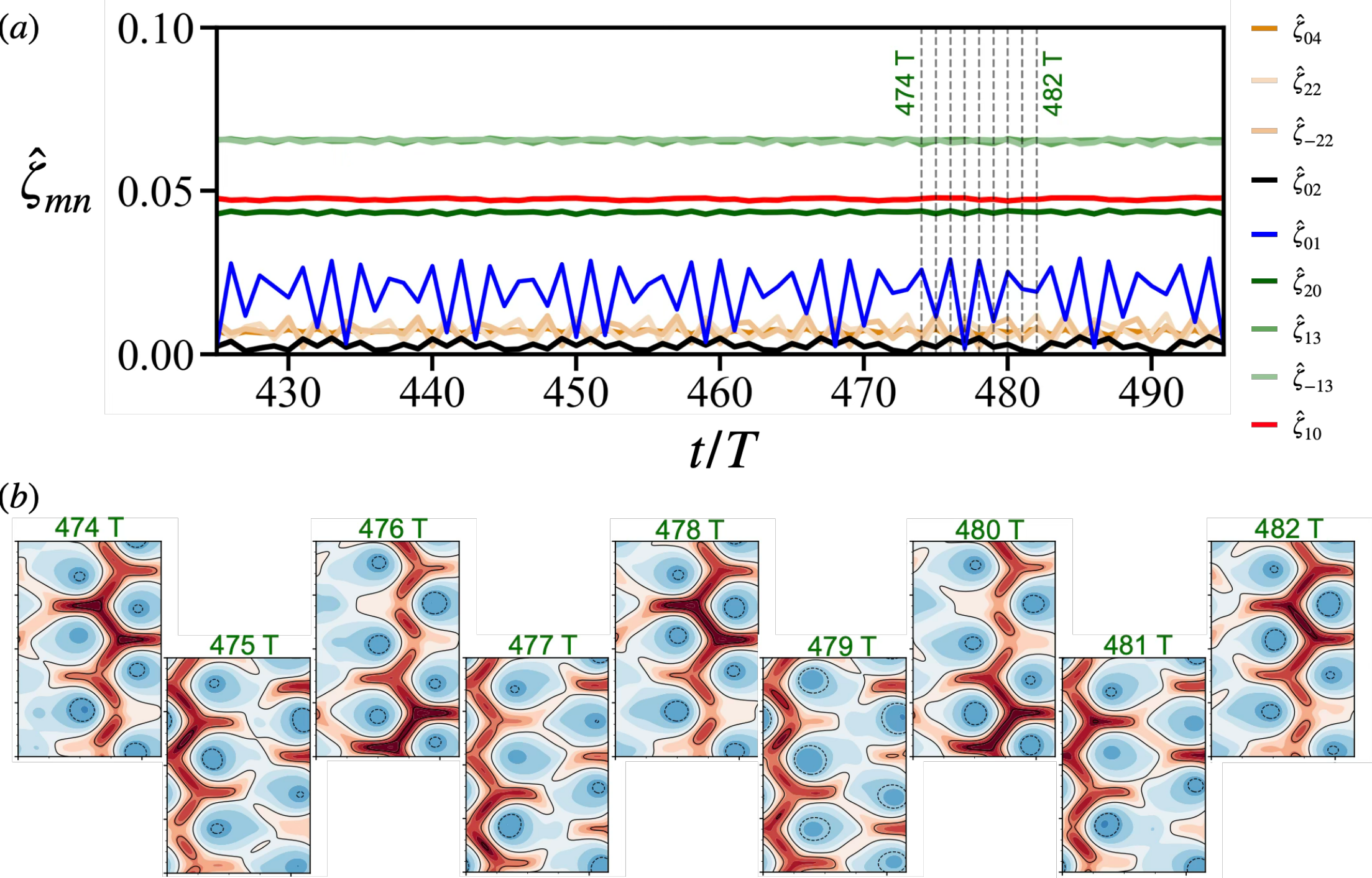}
    \caption{(a) Magnified view of the Fourier mode amplitudes over 
    %LSTAug26 Please check time range of (a), I changed it in the text by mistake
    $t \in [426T, 495T]$
    showing the oscillation of $\hat\zeta_{0,1}$ with a period that is between $8\,T$ and $9\,T$. The amplitudes of the other modes are nearly constant. (b) Consecutive interface contour snapshots from $t = 474\,T$ to $482\,T$. 
    The underlying backbone (colored in brown) alternates from left to right at each period $T$ (compare $475\,T$ to $476\,T$), corresponding to the subharmonic temporal behavior of $\hat\zeta_{1,0}$. The shift in intensity (dark brown color) along the backbone is related to the spatial phase of the doubly subharmonic mode $\hat\zeta_{0,1}$ and has approximate period $4\,T$ (compare $474\,T$ with $478\,T$).
    The figures span one full modulation cycle.}
    \label{fig:sss_3}
\end{figure*} 

\subsection{Dynamic SSS-I}
\label{sec:dyn}

Although the simulation appeared to be fairly steady at $\mathrm{SSS_s}$, we continued it to $t = 1150\,T$ to determine its long-term fate. 
There is no dramatic change in the pattern or in the spectrum after $340\,T$, but figure \ref{fig:sss_1}(b) shows that the pattern has shifted along the $y$ direction, and the backbone has become non-uniform. Figure \ref{fig:sss_2} shows the growth and rapid oscillations of certain modes.
More specifically, at $t \approx 350\,T$, the doubly subharmonic mode $\hat\zeta_{0,1}$ of wavenumber $k_c/4$ appears and develops sustained oscillations. This mode generates the
$4\lambda_c$-long modulation in $y$ visible in figure \ref{fig:sss_1}(b) for $t \geq 474$. The appearance of $\hat\zeta_{0,1}$ also breaks the $y$-reflection symmetry, as discussed in Appendix \ref{app:modif}. Figure \ref{fig:sss_2} also shows minor changes and oscillations in the modes comprising SSS$_s$.
There is also a resurgence of the primary hexagonal modes, but they remain weak and have little influence on the pattern. 
These changes all lead to a small but visible modulation of the kinetic energy envelope in figure \ref{fig:sss_1}(a).

Figure \ref{fig:sss_1}(b) for $t\geq 340\, T$ clearly displays the drift of the pattern in the $y$ direction, both in the backbone and in the shifting positions of the crests and troughs. 
Examination of the phases of the Fourier modes shows that the drift is related to the evolution of the phases $\phi_{1,3}$ and $\phi_{-1,3}$ of the secondary hexagonal modes, shown in figure \ref{fig:del_phi} and discussed in Appendix \ref{ss-A2}. 
These phases change slowly and at a constant rate, completing a %n entire 
cycle of $2\pi$ in approximately $120\,T$, after which the branches emanating from the backbone return to their initial locations. 

The short-time dependence of the superlattice regime is presented in figure \ref{fig:sss_3}. 
Figure \ref{fig:sss_3}(b) shows the evolution of the interface over $8\,T$. (Its most visible feature, already part of the basic SSS pattern and mentioned in \S~\ref{ss-short}, is that the backbone jumps every $T$ from the left half of the box to the right half. This behavior, caused by the subharmonic temporal behavior of the $x$-dependent mode $\hat\zeta_{1,0}$, is seen in figure \ref{fig-10} but not in figure \ref{fig:sss_1}(b) because for $t\geq 340\,T$, visualizations are presented only at even multiples of $T$.)
New features of the evolution of the backbones in figure \ref{fig:sss_3}(b) are due to the $y$-dependent mode $\tilde\zeta_{0,1}$, which appears only in this dynamic version of SSS$_s$. The overall intensity of the backbone is governed by the amplitude $\hat\zeta_{0,1}$; high-intensity moments like $476\,T$ and $478\,T$ coincide with the maxima in figure \ref{fig:sss_3}(a), and low-intensity moments like $477\,T$ coincide with the minima. The position in $y$ of the most intense regions is governed by the phase $\phi_{0,1}$ shown in figure \ref{fig:del_phi}. These intense regions travel intermittently along the $y$ axis, staying for $2\,T$ in the lower half of the box and then jumping to the upper half. 

The basic SSS-I state consists of four modes with two wavenumbers $\sqrt{3}k_c/2$ and $\sqrt{3}k_c/4$, and two temporal periods, $T$ and $2T$, respectively. Its dynamic version adds to this yet another significant mode, with a different wavenumber $k_c/4$, and three more timescales, all related to the motion of the position and intensity of the backbone. The resulting dynamical SSS-I is a state of great spatio-temporal complexity.

\section{Concluding remarks}
\label{sec_conclusion}

\begin{figure}
    \centering
    \includegraphics[width=\linewidth]
%    {Arbell_fig_6.pdf} 
    {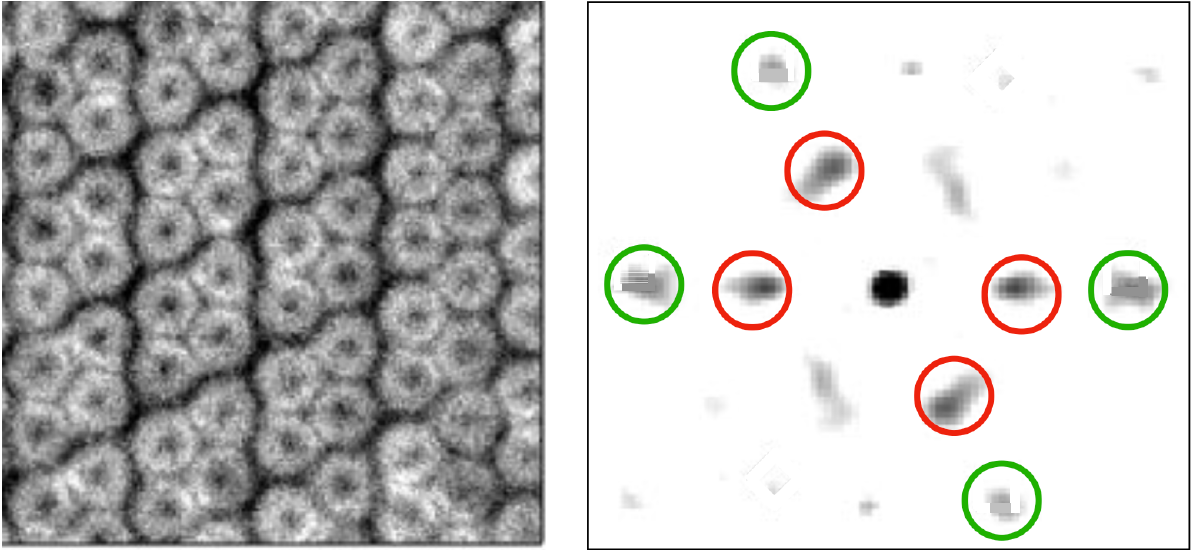} 
    \caption{Experimental SSS-I superlattice pattern, reproduced and adapted with permission from  Figure 6 of  \textcite{arbell2002pattern}. (a) Visualization of height via light intensity. (b) Spatial Fourier transform of (a), showing two hexagonal wave vector pairs (green circles, $k$) and two spatially subharmonic wave vector pairs (red circles, $k/2$). 
}
    \label{fig:Arbell_fig6}
\end{figure}

We have numerically simulated a pattern very much like the SSS-I pattern seen experimentally by \textcite{arbell2002pattern}.
At a forcing amplitude 25\% above the Faraday threshold and in a domain whose horizontal dimensions, 
%LSTAug15 domain size
$4\lambda_c/\sqrt{3}\times 4\lambda_c$ are chosen to accommodate hexagons. 
We have found a complex time-dependent state that resembles their SSS-I pattern.
Its Fourier spectrum is like that illustrated in figure \ref{fig:Arbell}: it consists of three secondary hexagonal modes with $k=\sqrt{3}k_c/2$ combined with the secondary subharmonic mode $\hat\zeta_{1,0}$ with $k=\sqrt{3}k_c/4$. The resulting interface forms a characteristic zig-zag configuration of crests and troughs with a $y$-oriented backbone or skeleton.

Moreover, we have obtained the SSS-I pattern via two quite different routes involving many intermediate states and modes. 
When initiated from the hexagonal state obtained at lower forcing acceleration $F_h$, the transition proceeds through
period doubling followed by an erratic transient involving eleven competing modes before terminating in the final state. 
In contrast, 
starting from a randomly perturbed low-amplitude 
hexagonal interface, a longer simulation leads to a transition through several fairly clearly-defined and long-lived states: 
the system evolves sequentially from a primary hexagonal pattern ${\rm Hex_p}$ to a
primary superlattice-like pattern ${\rm SSS_p}$, then to a secondary hexagonal state ${\rm Hex_s}$, and finally to the secondary superlattice state ${\rm SSS_s}$.
${\rm SSS_s}$ is eventually succeeded by a dynamic version of this superlattice state, in which the backbone is modulated in intensity and slowly drifts. 

For both the basic and dynamic versions of the superlattice states, we have been able to attribute their spatio-temporal features to their non-hexagonal spectral components, most notably the $x$-dependent $\tilde\zeta_{1,0}$ (perpendicular to the backbone) and the $y$-dependent $\tilde\zeta_{0,1}$ (parallel to the backbone).  The role of $\tilde\zeta_{2,-1}$, which plays a prominent role -- transient in Fig.\ \ref{fig:SL-C}, absent in Fig.\ \ref{fig:sss_2}, and permanent in Fig.\ \ref{fig:Spec_110}
 -- remains mysterious.

There are discrepancies between our SSS-I states and those of \textcite{arbell2002pattern}. The main difference is that the modes we observe have wavenumbers that differ from theirs by a factor of $\sqrt{3}/2$. 
We do not know the reason for this.
%Second, our spatially subharmonic wavevectors are present in only one direction, whereas two or three directions are also observed experimentally. The cause for these differences is probably domain size: our simulations were carried out in a small domain -- twice as large in each direction as the minimal hexagonal domain -- whereas the experiments of \textcite{arbell2002pattern} were carried out in a much larger domain. Convergence to SSS-I states both experimentally and numerically from two different initial conditions and in two different geometries nevertheless demonstrates the robustness of the basic SSS-I pattern. 
%
In addition, \textcite{arbell2002pattern} observed several variants of their SSS-I pattern involving spatially subharmonic wavevectors in more than one direction, indicated by the dotted arrows that we included in our Fig.\ \ref{fig:Arbell}. 
Figure 6 of \cite{arbell2002pattern}, reproduced in modified form in our Fig.\ \ref{fig:Arbell_fig6}, shows a pattern with subharmonic wavevectors in two directions, while their Figure 7 shows a state in which there is slow alternation between SSS-I states with subharmonic wavevectors in one, two, and three directions. 
They are necessarily absent from our simulations because, out of the secondary subharmonic modes ($k=\sqrt{3}k_c/4$) only $\tilde{\zeta}_{\pm 1,0}$ can exist in our computational domain. 
Other secondary subharmonic modes could, however, fit 
if we doubled the size of our domain. 
Alternatively, the full range of SSS-I patterns would fit in our domain if the hexagonal modes of our pattern had remained the primary ones, all of whose subharmonics ($k=k_c/2$) fit in our domain. 
It would be interesting to see if numerical simulations in a doubled domain would lead to these other versions of SSS-I.

%LSTAug18 Ask Alastair
The various versions of SSS-I are also explored and explained in terms of group theory and symmetry by \textcite{rucklidge2003secondary,Matthews_2004}, using temporally discrete (period $T$) versions of the governing equations. 
%These authors study only temporally discrete (period $T$) versions of the governing equations, thus disallowing dynamic states which pass through the three different versions. 
Within this framework, they state that the SSS-I with two subharmonic wave vectors is unstable, while either the version with one or with three subharmonic wave vectors may be stable.
%
%Conversely, halving the $y$ dimension of our domain would exclude two of the secondary hexagonal modes, thus suppressing the transition from Hex$_p$ to Hex$_s$. Halving the $y$ dimension would also suppress the instability to the dynamic SSS-I state. 

Finally, we mention that SSS-I is only one of the many patterns observed by \textcite{arbell2002pattern} for two-frequency-driven Faraday waves. These other patterns may also merit numerical investigations similar to this one.

\section*{Acknowledgements}
This work was supported by the Engineering and Physical Sciences Research Council, UK, through the  PREMIERE (EP/T000414/1) programme grant, the ANTENNA Prosperity Partnership grant (EP/V056891/1), and by the National Research Foundation of Korea(NRF) grant funded by the Korea government (MSIT) (No. RS-2025-02302984). O.K.M. acknowledges funding from PETRONAS and the Royal Academy of Engineering for a Research Chair in Multiphase Fluid Dynamics. D.P. and L.K. acknowledge HPC facilities provided by the Imperial College London Research Computing Service. D.P. acknowledges the Imperial College London President’s PhD scholarship. D.J. and J.C. acknowledge support through HPC/AI computing time at the Institut du Developpement et des Ressources en Informatique Scientifique (IDRIS) of the Centre National de la Recherche Scientifique (CNRS), coordinated by GENCI (Grand Equipement National de Calcul Intensif) grant 2026 A0202B06721.
 N.P. acknowledges partial support from FONDECYT Regular through grants 1241498 and 1241542.
\section*{Data availability}
The data that support the findings of this article are not
publicly available. The data are available from the authors
upon reasonable request.
%\clearpage
%\bibliography{refs}
\appendix

\section{Fourier analysis and symmetry breaking}
\label{sec:appendix-fourier}

In what follows, we will discuss the origin of the symmetry breaking of the DNA superlattice due to the presence of the mixed-parity Fourier mode.     
%
%\subsection{General symmetry condition for Fourier modes}
We first establish 
the general symmetry condition upon which we will assess the symmetry conditions on the superlattice state. All the results presented here are based on the property of symmetry under 
an operator 
$\gamma$, which can be a reflection,
%LSTAug 18 What is this?
%a point symmetry, 
a rotation, or a translation. If the height of the interface $\zeta$ is parameterized by 
$\xb = \left(x,y\right)^T$, it is invariant under $\gamma$ if 
\begin{equation}
\zeta(\gamma\xb) = \zeta(\xb),\: \forall \xb\in\mathbb{R}^2.
\label{eq:general_symmetry}
\end{equation}
If we apply \eqref{eq:general_symmetry} to the Fourier expansion of $\zeta$, 
\begin{align}
    \zeta(\mathbf{x}, t) &= \sum_{m,n} e^{i \mathbf{k}_{m,n} \cdot \mathbf{x}}      \tilde{\zeta}_{m,n}(t), \quad j=0, 1, \ldots 
    \label{eq:FT2}
\end{align}
we can transfer the action of $\gamma$ successively from $\xb$ to $\kb_{m,n}$ and then to $(m,n)$ to conclude that  \eqref{eq:general_symmetry} is equivalent to
\begin{equation}
\tilde{\zeta}_{\gamma^\prime(m,n)}=\tilde{\zeta}_{(m,n)},\:\forall(m,n)\in
\mathbb{Z}^2.
\label{eq:general_symmetry_Fourier_final}
\end{equation} 
for some other transformation $\gamma^\prime$.
From this condition, we extract relations between the modes $\tilde{\zeta}_{m,n}$ of the grid. 

\subsection{Hexagonal symmetry in the 
simulation domain}\label{ss-A1}
Here, we establish the conditions for a pattern to have a hexagonal symmetry in a rectangular box whose aspect ratio is $\sqrt{3}$. The simplest way to do so is to pass into the Fourier space. We first recall the elementary symmetries that belong to the hexagonal symmetry group $D_6$. 
This group comprises a reflection $\mu$ across any diagonal of the hexagon and a rotation of angle $\pi/3$, denoted $\rho_{\pi/3}$, plus all the possible compositions of these isometries; see \cite{golubitsky1984symmetries,hoyle2006pattern}. 
For simplicity, we assume the center of rotation to be the origin. 

\begin{figure}
    \centering
    \includegraphics[width = 0.65\linewidth]{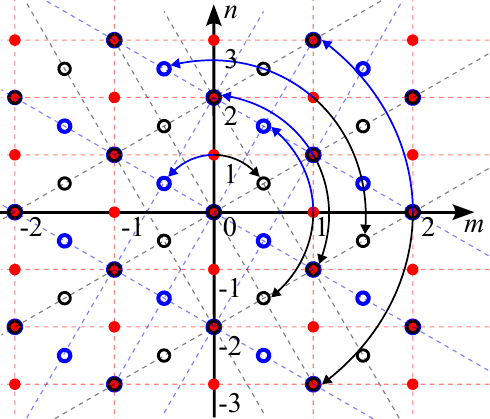}
    \caption{ Grids formed by the $\kb_{m,n}$ of the box modes (red disks) and their images by rotations of $\pi/3$ (blue circles) and $-\pi/3$ (black circles). The three grids coincide on points characterized by $m+n$ even
    where red, black and blue circles are superposed. Only the even modes have images on the red grid, as suggest the few examples of rotation showed by the blue and black bows of circles. Orange arrows represent two base vectors of the coarsest hexagonal network containing the whole red grid.}
    \label{fig-11_modes}
    \end{figure}
    
  \begin{figure*}
    \centering
    \includegraphics[width = 0.9\linewidth]{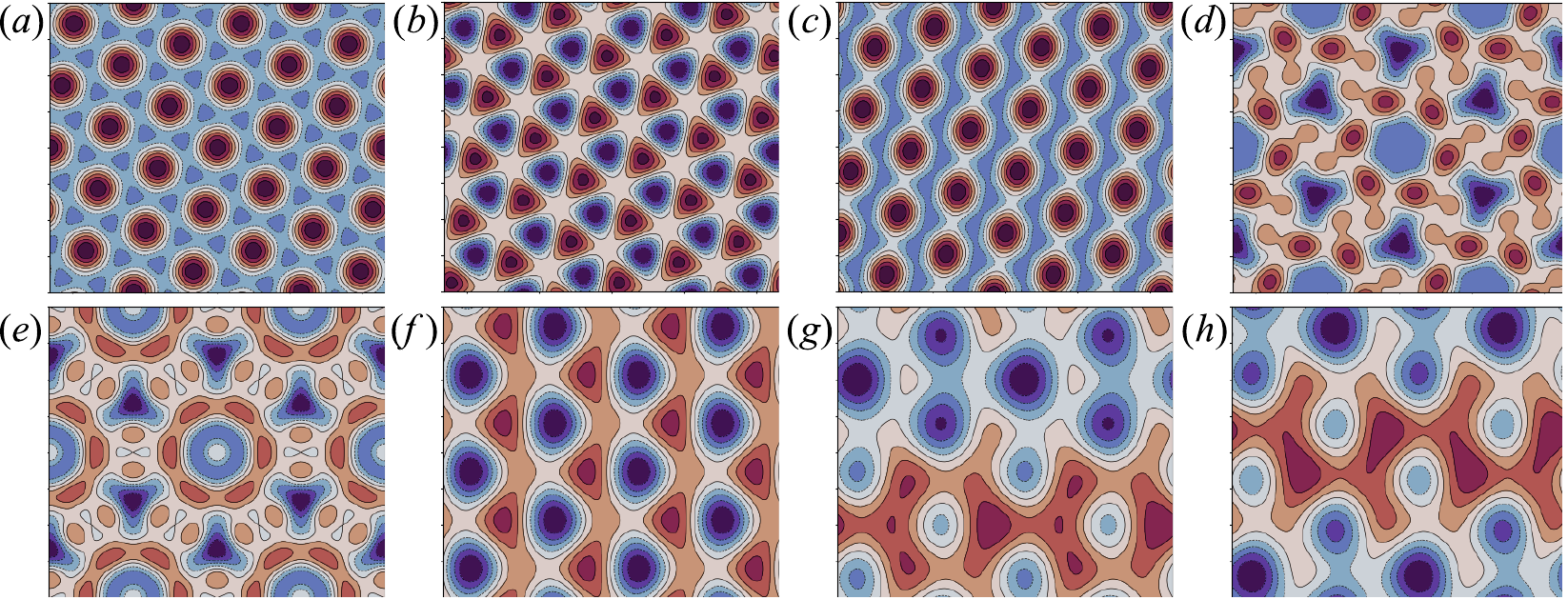}
    \caption{  
    (a)-(c) Examples of patterns formed by modes $\tilde{\zeta}_{3,-1}$, $\tilde{\zeta}_{2,4}$ and $\tilde{\zeta}_{-1,5}$. (d) pattern formed by two hexagons not aligned angularly: $\tilde{\zeta}_{4,2}$, $\tilde{\zeta}_{1.7}$ and $\tilde{\zeta}_{-3,5}$ and $\tilde{\zeta}_{2,0}$, $\tilde{\zeta}_{1,3}$ amd $\tilde{\zeta}_{-1,3}$. (e) Pattern formed by the same modes as in (e), plus the modes $\tilde{\zeta}_{4,-2}$, $\tilde{\zeta}_{1,-7}$ and $\tilde{\zeta}_{-3,-5}$, which restore the axial symmetry across the axis $y=0$. (f) Pattern formed by modes $\tilde{\zeta}_{2,0}$, $\tilde{\zeta}_{1,3}$ amd $\tilde{\zeta}_{-1,3}$. (g,h) Examples of patterns formed by modes $\tilde{\zeta}_{2,0}$, $\tilde{\zeta}_{1,3}$, $\tilde{\zeta}_{-1,3}$ and $\tilde{\zeta}_{0,1}$ with and without symmetry across a horizontal axis.
    }
    \label{fig-11_patterns}
\end{figure*}

We first wish to determine the constraints on gridpoints $(m,n)$ such that their images under rotation are contained in the grid. Applying the rotation $\rho_{-\pi/3}$ to $(m,n)$ yields:
\begin{equation}
   \rho_{-\frac{\pi}{3}}\left(
   \begin{array}{c}
   \vs m\\
   n
   \end{array}\right)
   = \left(\begin{array}{c}
   \vs (m+n)/2\\
   (-3m+n)/2
   \end{array}\right).
   \label{eq:pi3trans}
\end{equation}
Definition \eqref{eq:pi3trans} provides an example of how a symmetry operator $\gamma$ on the spatial points $(x,y)$ is used to define $\gamma^\prime$ on Fourier indices $(m,n)$, as mentioned in  \eqref{eq:general_symmetry_Fourier_final}.
In order for \eqref{eq:pi3trans} to be on the grid, $m+n$ and $-3m+n$ must be even, 
meaning that $m$ and $n$ must be of the same parity.
Figure \ref{fig-11_modes} shows that the application of $\rho_{\pi/3}$ sends 
modes with $m+n$ odd, i..e.\ with $m$ and $n$ of opposite parity,
to images with non-integer indices that do not belong to the rectangular grid.
In order for a pattern to have hexagonal symmetry, it is not sufficient for its Fourier coefficients to remain in the grid under rotation by $\pi/3$ and all its multiples, they must also be invariant:
\begin{equation}
\begin{array}{rl}
\vs\tilde\zeta_{m,n} &= \tilde\zeta_{\frac{m+n}{2},\frac{-3m+n}{2}} = \tilde\zeta_{\frac{-m+n}{2},\frac{-3m-n}{2}} = \tilde\zeta_{-m,-n}\\&= \tilde\zeta_{\frac{-m-n}{2},\frac{3m-n}{2}} = \tilde\zeta_{\frac{m-n}{2},\frac{3m+n}{2}}.%,\:\forall(m,n)\in\mathbb{Z}^2.
\end{array}
\label{eq:rotation_inv}
\end{equation}

%Substituting \eqref{eq:pi3trans} into \eqref{eq:general_symmetry_Fourier_final} leads to the symmetry condition under compositions of rotations by $-\pi/3$ about (0,0):
%
%As a consequence of \eqref{eq:rotation_inv}, the symmetry under $\rho_{\pi/3}$ is systematically broken, as well as the hexagonal symmetry, if a pattern contains a mode $(m,n)$ such that $m+n$ is odd (i.e. $m$, $n$ are of different parities).

It follows from \eqref{eq:rotation_inv} combined with the condition $\tilde{\zeta}_{-m.-n}=\tilde{\zeta}_{m.n}$ that all $\tilde{\zeta}_{m,n}$ are real when the 
pattern has hexagonal symmetry with center of rotation (0,0). In particular, 
\begin{equation}
\phi_{m,n} - \phi_{\frac{m+n}{2},\frac{-3m+n}{2}} + \phi_{\frac{-m+n}{2},\frac{-3m-n}{2}} = 0\mod(\pi)
\label{eq:args}
\end{equation} 
where $\phi_{m,n} = \arg(\tilde{\zeta}_{m,n})$
The motivation for writing 
\eqref{eq:args} in this way 
is that it also holds in the more general case in which the center of rotation is not at the origin but at a location $\xb_0$, so that $\tilde{\zeta}_{m,n}$ are not all real. 
Equations \eqref{eq:rotation_inv} and \eqref{eq:args} then apply in a system of coordinates translated by $-\xb_0$. After translating \eqref{eq:args} back to the original coordinates, we obtain:
\begin{equation}
\begin{array}{c}
\vs\phi_{m,n} - \phi_{\frac{m+n}{2},\frac{-3m+n}{2}} + \phi_{\frac{-m+n}{2},\frac{-3m-n}{2}}\\
+ \left(\kb_{m,n}-\kb_{\frac{m+n}{2},\frac{-3m+n}{2}}+\kb_{\frac{-m+n}{2},\frac{-3m-n}{2}}\right)\cdot\xb_0 = 0.
\end{array}
\label{eq:args2}
\end{equation}
The term in parentheses vanishes, so we again obtain \eqref{eq:args}.
If modes distributed on different hexagons cohabit, the hexagonal symmetry 
requires that the symmetry axes of the plane waves composing the pattern all intersect at the same point,
which fixes all of the $\phi_{m,n}$.

%LSTJun16 Do we need to mention this at all?
%A grid that contains all the modes of the box and their images by $\rho_{\pi/3}$ is that spanned by the vectors $\kb_{\frac{1}{2}\pm\frac{1}{2}} = \left(\sqrt 3  \mathbf e_x \pm ~\mathbf e_y\right)k_c/8$. This grid is the same as in \cite{golubitsky1984symmetries} and is defined by $\kb_{m/2,n/2}$ where $m+n$ is even. None of $\kb_{\frac{1}{2}\pm\frac{1}{2}}$ belong to the red grid of figure \ref{fig-11_modes}; they cannot appear in the simulations since they are incompatible with the periodic conditions of the domain. 

%We first note that the application of $\rho_{\pi/3}$ sends some modes to images that are not modes of the box. Figure \ref{fig-11_modes} describes this phenomenon.

The hexagonal symmetry group also contains reflections.
The simplest case is reflection about the horizontal axis $y=0$, so that $\gamma (x,y)= (x,-y)$ and $\gamma^\prime(m,n)=(m,-n)$. A slightly more complicated case 
%The simplest cases are the horizontal or vertical reflections (possibly composed with rotations through $\pi/3$ and its multiples). 
is reflection across horizontal axis $y = y_0$. The pattern $\zeta(\xb)$ is invariant under reflection in $y = y_0$ if 
\begin{align}
\zeta(x,y) &= \zeta(x,2y_0-y),\:\forall(x,y)\in\mathbb{R}^2.
\label{eq:reflection_general}\\
    \tilde{\zeta}_{m,n}&=\tilde{\zeta}_{m,-n}\ed^{\id2\kb_{m,-n}\cdot(0,y_0)^T},\:\forall(m,n)\in\mathbb{Z}^2.
    \label{eq:reflection_C}
\end{align}
where, as in the case of rotation, the formalism has been modified to include a translation, here  $y\rightarrow y+y_0$. 
The equality \eqref{eq:reflection_C} between complex quantities can be decomposed into equations for the norms $\hat\zeta_{m,n}$ and the phases $\phi_{m,n}$
\begin{subequations}
\label{eq:reflection_RPhi}
\begin{align}
\vs\hat{\zeta}_{m,n} &= \hat{\zeta}_{m,-n}, \label{eq:reflection_RPhi_amp}\\
\vs\phi_{m,n}-\phi_{m,-n} &= 2\kb_{m,-n}\cdot(0,y_0)^T \mod 2\pi
\label{eq:reflection_RPhi_phase}
\end{align}
\end{subequations}
Note that \eqref{eq:reflection_RPhi_phase} does not fix an absolute phase for a given mode: if the phases of $\phi_{m,n}$ and $\phi_{m,-n}$ vary together, the pattern can conserve reflection symmetry. The modes $\tilde{\zeta}_{m,0}$ are completely free, because they are homogeneous in $y$, and hence 
admit any line $y=y_0$ as a symmetry axis. 

When all equations \eqref{eq:rotation_inv}, \eqref{eq:reflection_RPhi_amp} and \eqref{eq:reflection_RPhi_phase} are satisfied, we obtain the hexagons of Fig.\ \ref{fig-11_patterns}(a).
If the phase condition \eqref{eq:args} is not satisfied, 
%the hexagons lose $D_6$ symmetry in favor of $D_3$,
the symmetry group $D_6$ is replaced by $D_3$,
which is the symmetry group of equilateral triangles, as shown in Fig.\ \ref{fig-11_patterns}(b). If the norm condition based on \eqref{eq:rotation_inv} fails, we can obtain centered rectangles as shown in Fig.\ \ref{fig-11_patterns}(c). If the norm condition \eqref{eq:reflection_RPhi_amp} fails, asymmetric hexagons (symmetry group $Z_6$) as depicted in Fig.\ \ref{fig-11_patterns}(d) can be obtained. This pattern is composed of three modes on two distinct hexagons: $\tilde{\zeta}_{4,2}$, $\tilde{\zeta}_{1,7}$ and $\tilde{\zeta}_{-3,5}$ (norms $\hat\zeta_{m,n} = 0.5$, phases $\phi_{m,n} = 0$) and $\tilde{\zeta}_{2,0}$, $\tilde{\zeta}_{1,3}$ and $\tilde{\zeta}_{-1,3}$ (norms $\hat\zeta_{m,n} = 1$, phases $\phi_{m,n} = \pi$). Within the same hexagon, the norms are all equal and the phase condition is satisfied, but the symmetry axis of the first hexagon is not a horizontal line, which breaks the (horizontal) reflection symmetry. This symmetry is reestablished if the reflection of the first hexagon across the axis $m=0$ is added, namely the modes 
$\tilde{\zeta}_{4,-2}$, $\tilde{\zeta}_{3,5}$ and $\tilde{\zeta}_{-1,7}$, all with norms $\hat\zeta_{m,n} = 0.5$ and the appropriate phases $\phi_{m,n} = 0$, 
as has been done to produce Fig.\ \ref{fig-11_patterns}(e).

In summary, a pattern has horizontal reflection symmetry if it contains only modes whose indices have the same parity, the norms of the modes are the same within a hexagon, and its phases satisfy \eqref{eq:args}. Equations \eqref{eq:reflection_RPhi} can be generalized to other directions of reflection axes. 

\subsection{Phase condition for horizontal reflection symmetry of a superposition of 
\texorpdfstring{$\tilde{\zeta}_{2,0}$}{zeta-2,0},
\texorpdfstring{$\tilde{\zeta}_{1,3}$}{zeta-1,3} and
\texorpdfstring{$\tilde{\zeta}_{-1,3}$}{zeta-minus-1,3}}
\label{ss-A2}
We consider the three hexagonal modes
$\tilde{\zeta}_{2,0}$, $\tilde{\zeta}_{1,3}$ and $\tilde{\zeta}_{-1,3}$, associated with the wave vectors $\kb_{2,0}$, $\kb_{1,3}$ and $\kb_{-1,3}$, respectively, where:

\begin{equation}
\kb_{2,0} = \frac{k_c}{4}\left(2\sqrt{3},0\right),\: \kb_{1,3} = \frac{k_c}{4}\left(\sqrt{3},3\right),\:  \kb_{-1,3} = \frac{k_c}{4}\left(-\sqrt{3},3\right).
\label{eq:k}
\end{equation}

The interface height $\zeta$ is composed only of 
these three modes and their complex conjugates.
We seek the condition for the patterns formed by the free surface to be reflection-symmetric about 
$y=y_0$. In particular, we are interested in the case 
in which all three modes have the same amplitude. Then, we set:
\begin{equation}
\tilde{\zeta}_{2,0} = Z\ed^{\id\phi_{2,0}},\: \tilde{\zeta}_{1,3} = Z\ed^{\id\phi_{1,3}},\: \tilde{\zeta}_{-1,3} = Z\ed^{\id\phi_{-1,3}}.
\label{eq:zeta_modes}
\end{equation}
%LSTJun16 not necessary.
%with $Z$ a positive real quantity and $\phi_{lm}$ real phases. 
We would like to know if there is always such an axis for any triplet $\{\phi_{2,0},\phi_{1,3},\phi_{-1,3}\}\in[0,2\pi]^3$. \\
A pattern that is reflection-symmetric about
$y=y_0$ satisfies \eqref{eq:reflection_RPhi}. We recall that there is no constraint on $\tilde{\zeta}_{2,0}$ because it is homogeneous in $y$. Substituting \eqref{eq:k}-\eqref{eq:zeta_modes} into \eqref{eq:reflection_RPhi_phase} for $(m,n) = (1,3)$ and replacing $-\phi_{m,-n}$ by $\phi_{-m,n}$ yields: 
\begin{equation}
2\pi\frac{y_0}{4\lambda_c}+\frac{\phi_{1,3}+\phi_{-1,3}}{6} = 0 \mod \frac{\pi}{3}.
\label{eq:symmetry_condition}
\end{equation}
The relation \eqref{eq:symmetry_condition} 
establishes the dependence of $y_0$ on $\phi_{1,3}$ and $\phi_{-1,3}$. 
As a result, there exists a line $y=y_0$ about which $\zeta$ has a reflection symmetry for any pair $\left\{\phi_{1,3},\phi_{-1,3}\right\}$, provided that $\tilde{\zeta}_{1,3}$ and $\tilde{\zeta}_{-1,3}$ have the same amplitude $Z$. Since $\tilde{\zeta}_{2,0}$ is homogeneous in $y$, equation (\ref{eq:symmetry_condition}) would remain unchanged if $\phi_{2,0}$ or even the norm $\hat{\zeta}_{2,0}$ were varied. In Fig.\ \ref{fig-11_patterns}(f), we have plotted an example of such a pattern, with arbitrary phases. If $\phi_{1,3}+\phi_{-1,3}$ were varied, the horizontal symmetry axis would remain but would be translated vertically. 

In conclusion, the temporal variation of $\phi_{-1,3}$ and $\phi_{1,3}$ is not responsible for the breaking of the pattern symmetry across a horizontal axis, but merely 
translates the axis of symmetry.

\subsection{Modification of the phase condition when 
\texorpdfstring{$\tilde{\zeta}_{0,1}$}{zeta-01}
is added}\label{ss-A3}
\label{app:modif}
We add to $\zeta$ of the previous section the mode $\tilde{\zeta}_{0,1} = Z'\ed^{\id\phi_{0,1}}$ associated with the wave vector $\kb_{0,1} = \frac{k_c}{4}\left(0,1\right)$.
Now \eqref{eq:reflection_RPhi_phase} generates two conditions that must be satisfied simultaneously:
\begin{equation}
\left\{
\begin{array}{l}
\vs\displaystyle{2\pi\frac{y_0}{4\lambda_c}+\frac{\phi_{1,3}+\phi_{-1,3}}{6} = 0 \mod \frac{\pi}{3}}\\
\displaystyle{2\pi\frac{y_0}{4\lambda_c}+\phi_{0,1} = 0 \mod \pi}
\end{array}
\right.
\label{eq:symmetry_condition_01_1}
\end{equation}
By combining these two conditions, we deduce that a horizontal symmetry axis exists only when:
\begin{equation}
\frac{\phi_{1,3}+\phi_{-1,3}}{6}-\phi_{0,1} = 0 \mod \frac{\pi}{3}
\label{eq:symmetry_condition_01_2}
\end{equation}
The equation of the axis is given by either of the two conditions of (\ref{eq:symmetry_condition_01_1}). Figures \ref{fig-11_patterns}(g,h) show that some combinations of phases $\left\{\phi_{1,3},\phi_{-1,3},\phi_{0,1}\right\}$ break the reflection symmetry in $y=y_0$.
In Fig.\ \ref{fig-11_patterns}(g), a case satisfying the condition (\ref{eq:symmetry_condition_01_2}) has been chosen ($\left\{\phi_{1,3},\phi_{-1,3},\phi_{0,1}\right\} = \left\{\frac{4\pi}{3},\frac{5\pi}{3}, \frac{\pi}{2}\right\}$). If, for example, $\phi_{0,1}$ were increased by a multiple of $\pi/3$, we would still observe horizontal reflection symmetry, but about a different $y$. In Fig.\ \ref{fig-11_patterns}(h), $\left\{\phi_{1,3},\phi_{-1,3},\phi_{0,1}\right\} = \left\{\frac{4\pi}{3},\frac{5\pi}{3}, 0\right\}$ does not satisfy (\ref{eq:symmetry_condition_01_2}), hence $\zeta$ no longer has a horizontal reflection symmetry axis.

To conclude, when the mode $\tilde{\zeta}_{0,1}$ is present, not all quadruplets $\left\{\phi_{2,0},\phi_{1,3},\phi_{-1,3},\phi_{0,1}\right\}$ 
are horizontally reflection-symmetric.
The addition of $\tilde{\zeta}_{0,1}$ with an arbitrary phase is likely responsible for the breaking of reflection symmetry observed at time $t = 474\,T$ and thereafter; 
see section \ref{ss_DNA}, Fig.\ \ref{fig:sss_1}(b)). 
This is illustrated in Fig.\ \ref{fig:del_phi}, which tracks the phases $\phi_{0,1}$, $\phi_{1,3}$, and $\phi_{-1,3}$. 
As shown in Fig.\ \ref{fig:del_phi}, the phases $\phi_{1,3}$ and $\phi_{-1,3}$ are out of phase by $\pi/2$, while $\phi_{0,1}$ oscillates erratically. This phase misalignment results in a deviation $\Delta \phi$, as depicted in Fig.\ \ref{fig:del_phi}(b).  
%\clearpage

\section{Pattern evolution for \texorpdfstring{$F_h < F < 1.25\, F_h$}{flo}}
\label{sec:lowerF}

\begin{figure*}
\centering
\includegraphics[width=0.9\linewidth]{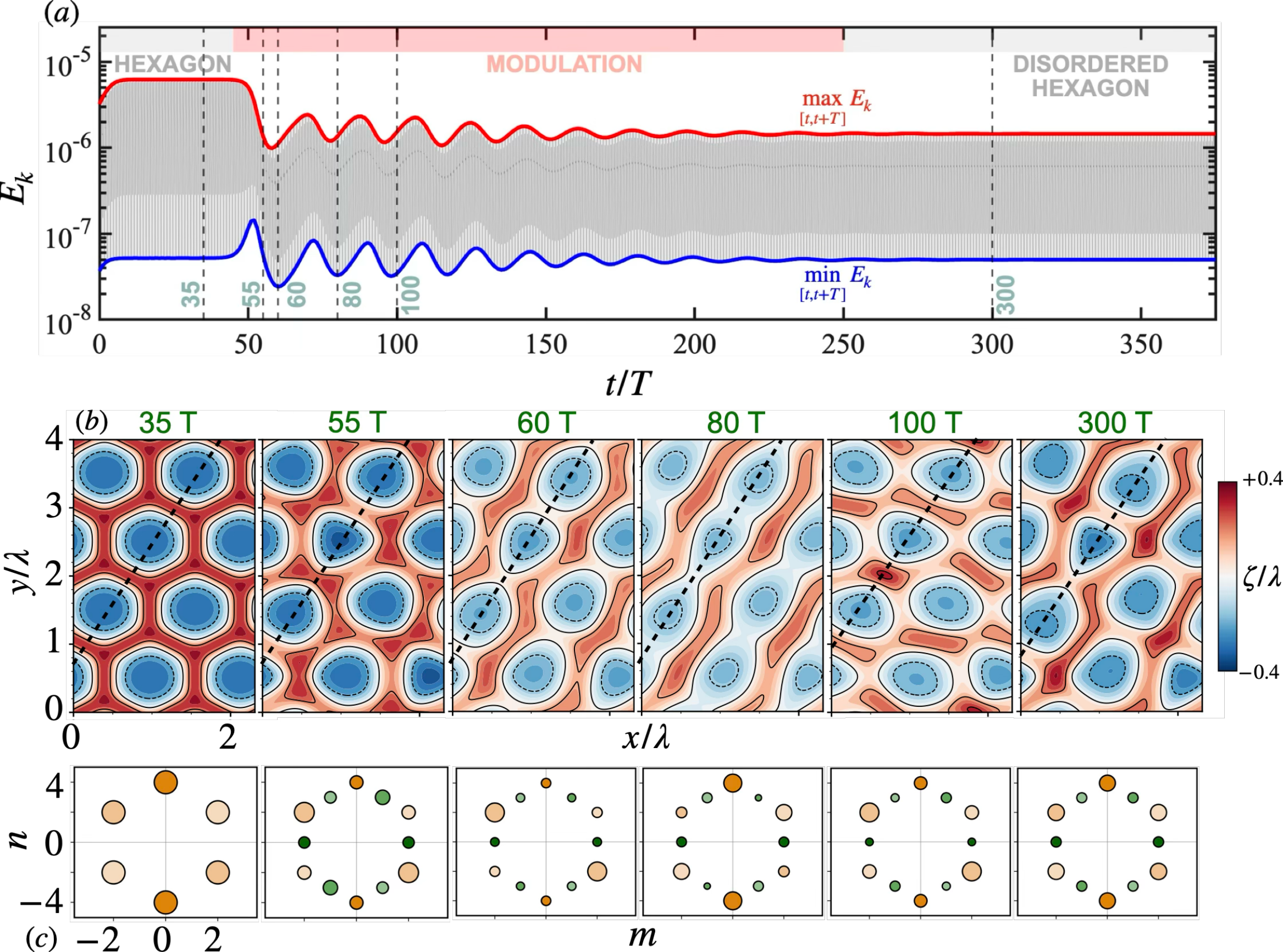}
\caption{Pattern evolution for $F=1.05\,F_h$. (a)  
    The temporal evolution of the kinetic energy, where the red and blue envelopes trace the maximum and minimum kinetic energy over a window averaged over $T$. The temporal evolution is decomposed into three stages: hexagons, modulation, and disordered hexagons. Dashed lines show axis of reflection symmetry inherited from hexagonal lattice.
    (b)  Contour plots (top) and Fourier spectra (bottom) of the interface height at instants indicated by vertical dotted lines in (a). 
    Color codes of Fourier modes as in Fig.\ \ref{fig:SL-A}.
  The size of the markers is proportional to their amplitudes.
      }
      \label{fig:Vis_105}
\end{figure*}

\begin{figure*}
\centering
\includegraphics[width=0.9\linewidth]{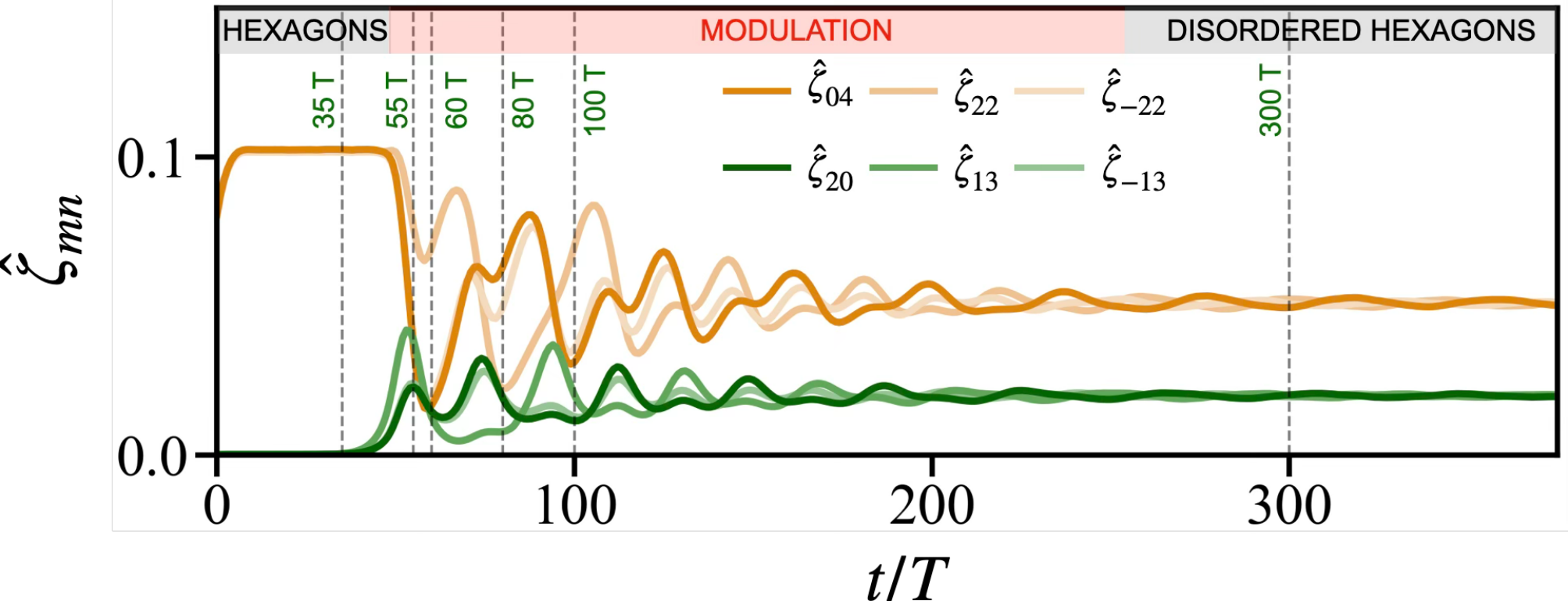}
\caption{Temporal evolution for $F=1.05\,F_h$ of the Fourier mode amplitudes 
    for the primary (orange) and secondary (green) hexagonal modes.
    The color coding matches that of the Fourier space grid in  Fig.\ \ref{fig:SL-A}(a).
Vertical dotted lines correspond to the instants shown in Fig.\ \ref{fig:Vis_105}.
    }
    \label{fig:Spec_105}
\end{figure*}

Figures \ref{fig:Vis_105} and \ref{fig:Spec_105} show the evolution at $F=1.05\, F_h$ starting from the fully developed hexagons at $F=F_h$. Fig.\ \ref{fig:Vis_105}(a,b) show that the hexagonal patttern persists until about $45\,T$, when oscillations appear.  The three primary hexagonal mode amplitudes separate from one another while the secondary modes appear, as shown in Fig.\ \ref{fig:Spec_105}.
The oscillations decay and the final pattern, seen in Fig.\ \ref{fig:Vis_105}(b) at $300\,T$, is steady.
%resembling asymmetric hexagons.
Fig.\ \ref{fig:Vis_105}(b) shows that one prominent tilted axis of reflection symmetry remains, inherited from the hexagonal pattern. 
The amplitudes of the sets of primary and secondary hexagonal modes in Fig. \ref{fig:Vis_105}(c) and \ref{fig:Spec_105} are equal within each set, so there must be hexagonal asymmetry in the phases, i.e.\ \eqref{eq:args} is violated.

\begin{figure*}
\centering
\includegraphics[width=0.9\linewidth]{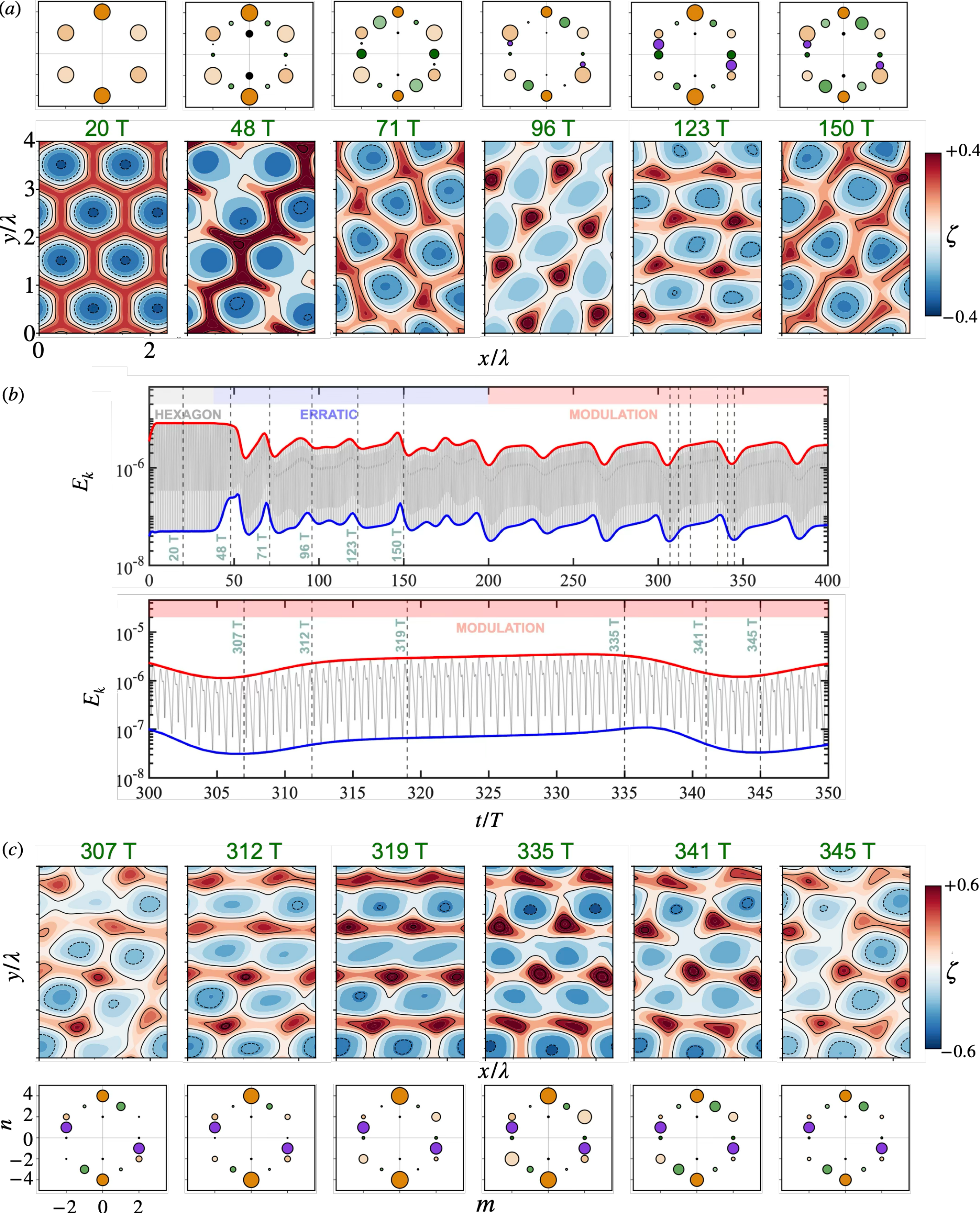}
\caption{Pattern evolution for $F=1.10\,F_h$, divided into three phases: hexagonal, erratic, and modulation.
 (a)  Contour plots of the interface height (top) and Fourier spectra (bottom) over the hexagonal and erratic phases at instants indicated by vertical dotted lines in (b). Temporal evolution of the kinetic energy, with $T$-averaged envelope of maximum and minimum values over the entire evolution (top) and over a modulation period (bottom).
(c)  Contour plots of the interface height (top) and Fourier spectra (bottom) over one modulation period at instants indicated by vertical dotted lines in (b).
Color codes of Fourier modes as in Fig.\ \ref{fig:SL-A}.
  The size of the markers is proportional to their amplitudes.}
\label{fig:Vis_110}
\end{figure*}

\begin{figure*}
\centering
\includegraphics[width=0.85\linewidth]{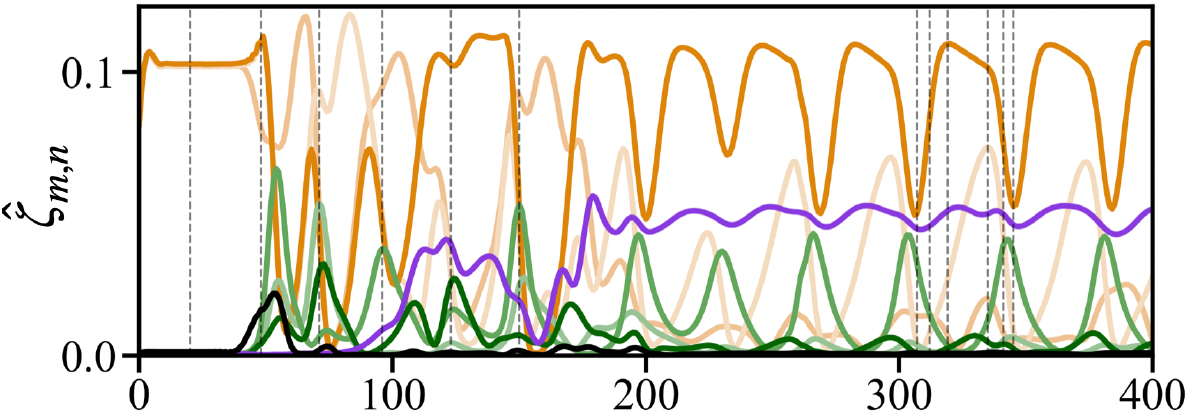}
\caption{Temporal evolution for $F=1.10\,F_h$ of the Fourier mode amplitudes for the primary (orange), secondary hexagonal modes (green), and the anomalous mode $\hat\zeta_{-2,1}$ (purple).
The color coding matches that of the Fourier space grid in  Fig.\ \ref{fig:SL-A}(a). Vertical dotted lines correspond to the instants shown in Fig.\ \ref{fig:Vis_110}. }
    \label{fig:Spec_110}
\end{figure*}

Figures \ref{fig:Vis_110} and \ref{fig:Spec_110} show the evolution at $F=1.10\, F_h$, again from the fully developed hexagons at $F=F_h$. Figure \ref{fig:Vis_110}(a,b) show that the hexagonal regime is succeeded by an erratic transient. The state at $150\,T$, near the end of the erratic period,
looks very similar to the final state for $1.05\,F_h$  at $300\,T$. A regime which is periodic in the energy and principal modes with period $38\,T$ is established at around $250\,T$. Figure \ref{fig:Vis_110}(c) shows the pattern during one such period, where we can note that the pattern at 
$345\,T$ is translated in both $x$ and $y$ with respect to that at $307\,T$ and that the states at $312\,T$ and $319\,T$  resemble the beaded stripes observed in \cite{perinet2012alternating}.
Figures \ref{fig:Vis_110}(c) and \ref{fig:Spec_110} show that the modes involved in the asymptotic periodic state are two of the primary hexagonal modes, one secondary hexagonal mode, and the anomalous mode $\zeta_{-2,1}$ which was large but transient in one of the superlattice cases at $1.25\,F_h$; see Fig.\ 
\ref{fig:SL-C}. The behavior of the anomalous and the hexagonal modes with smaller amplitudes deviate from exact periodicity.

%\clearpage
%\subsection{Evolution for $F=1.10\,F_h$}
%\clearpage
\bibliography{refs}

\end{document}